\documentclass[preprint,superscriptaddress]{revtex4-2}

\usepackage[labelfont=bf]{caption}
\usepackage{times}
\usepackage{graphicx}
\usepackage{amsmath}
\usepackage{amssymb}
\usepackage[colorlinks,linkcolor=blue,citecolor=blue,urlcolor=blue,hyperindex,pdfstartview=FitH,plainpages=false]{hyperref}
\usepackage{float}

\newcommand {\TS}{TiSe$_2$}
\newcommand {\mJcm}{mJ/cm$^2$}

\begin{document}

\title{Optical manipulation of electronic dimensionality in a quantum material}

\author{Shaofeng Duan}
\thanks{These two authors contributed equally}
\affiliation{Key Laboratory of Artificial Structures and Quantum Control (Ministry of Education), Shenyang National Laboratory for Materials Science, School of Physics and Astronomy, Shanghai Jiao Tong University, Shanghai 200240, China}
\author{Yun Cheng}
\thanks{These two authors contributed equally}
\affiliation{Key Laboratory for Laser Plasmas (Ministry of Education), School of Physics and Astronomy, Shanghai Jiao Tong University, Shanghai 200240, China}
\author{Wei Xia}
\affiliation{School of Physical Science and Technology, ShanghaiTech University, Shanghai 201210, China}
\author{Yuanyuan Yang}
\affiliation{Key Laboratory of Artificial Structures and Quantum Control (Ministry of Education), Shenyang National Laboratory for Materials Science, School of Physics and Astronomy, Shanghai Jiao Tong University, Shanghai 200240, China}
\author{Chengyang Xu}
\affiliation{Key Laboratory of Artificial Structures and Quantum Control (Ministry of Education), School of Physics and Astronomy, Shanghai Jiao Tong University, Shanghai 200240, China}
\author{Fengfeng Qi}
\affiliation{Key Laboratory for Laser Plasmas (Ministry of Education), School of Physics and Astronomy, Shanghai Jiao Tong University, Shanghai 200240, China}
\author{Chaozhi Huang}
\author{Tianwei Tang}
\affiliation{Key Laboratory of Artificial Structures and Quantum Control (Ministry of Education), Shenyang National Laboratory for Materials Science, School of Physics and Astronomy, Shanghai Jiao Tong University, Shanghai 200240, China}
\author{Yanfeng Guo}
\affiliation{School of Physical Science and Technology, ShanghaiTech University, Shanghai 201210, China}
\author{Weidong Luo}
\affiliation{Key Laboratory of Artificial Structures and Quantum Control (Ministry of Education), School of Physics and Astronomy, Shanghai Jiao Tong University, Shanghai 200240, China}
\affiliation{Institute of Natural Sciences, Shanghai Jiao Tong University, Shanghai 200240, China}
\author{Dong Qian}
\affiliation{Key Laboratory of Artificial Structures and Quantum Control (Ministry of Education), Shenyang National Laboratory for Materials Science, School of Physics and Astronomy, Shanghai Jiao Tong University, Shanghai 200240, China}
\author{Dao Xiang}
\email{dxiang@sjtu.edu.cn}
\affiliation{Key Laboratory for Laser Plasmas (Ministry of Education), School of Physics and Astronomy, Shanghai Jiao Tong University, Shanghai 200240, China}
\affiliation{Zhangjiang Institute for Advanced Study, Shanghai Jiao Tong University, Shanghai 200240, China}
\affiliation{Tsung-Dao Lee Institute, Shanghai Jiao Tong University, Shanghai 200240, China}
\author{Jie Zhang}
\affiliation{Key Laboratory for Laser Plasmas (Ministry of Education), School of Physics and Astronomy, Shanghai Jiao Tong University, Shanghai 200240, China}
\author{Wentao Zhang}
\email{wentaozhang@sjtu.edu.cn}
\affiliation{Key Laboratory of Artificial Structures and Quantum Control (Ministry of Education), Shenyang National Laboratory for Materials Science, School of Physics and Astronomy, Shanghai Jiao Tong University, Shanghai 200240, China}

\begin{abstract}
\newpage
\section*{Abstract}

Exotic phenomenon can be achieved in quantum materials by confining electronic states into two dimensions. For example, relativistic fermions are realised in a single layer of carbon atoms \cite{CastroNeto2009}, the quantized Hall effect can result from two-dimensional (2D) systems \cite{Klitzing1980,Tsui1982}, and the superconducting transition temperature can be enhanced significantly in a one-atomic-layer material \cite{He2013,Tan2013}. Ordinarily, 2D electronic system can be obtained by exfoliating the layered materials, growing monolayer materials on substrates, or establishing interfaces between different materials. Herein, we use femtosecond infrared laser pulses to invert the periodic lattice distortion sectionally in a three-dimensional (3D) charge density wave material (1T-\TS), creating macroscopic domain walls of transient 2D ordered electronic states with exotic properties. The corresponding ultrafast electronic and lattice dynamics are captured by time- and angle-resolved photoemission spectroscopy \cite{Yang2019} and MeV ultrafast electron diffraction \cite{Qi2020}. Surprisingly, a novel phase with enhanced density of states and signatures of potential energy gap opening near the Fermi energy, is identified in the photoinduced 2D domain wall near the surface. Such optical modulation of atomic motion is a new path to realise 2D electronic states and will be a new platform for creating novel phases in quantum materials.

\end{abstract}

\maketitle
%\clearpage

\section*{Main}
The control of macroscopic quantum phenomena is the key to exploring quantum materials for applications. Ultrafast optical irradiation is a novel approach for manipulating quantum phases or even promoting the emergence of novel quantum phases, such as light-induced superconductivity \cite{Fausti2011,Mitrano2016}, optical switching to hidden phases \cite{Ichikawa2011,Stojchevska2014}, coherent control of phases \cite{Rini2007,Li2019,Sie2019,Horstmann2020}, ultrafast photon-induced metastable states \cite{Morrison2014,Nova2019}, and light-modulated exotic phases \cite{Wang2013}. In particular, photons coherently coupled to the lattice vibrations or phase order parameter can largely drive the excitations far from equilibrium, which could be an effective method to realise novel phases in a material. However, the possibility of realising novel phases by optically manipulating electronic dimensionality is still a largely open subject.

Ultrafast photon excitation in a charge density wave (CDW) material can induce a complete inversion of the phase order via an electron--phonon scattering process, while the symmetry of the inverted phase ($\Phi=1$) is equivalent to the original phase ($\Phi=-1$) \cite{Lian2020}. Macroscopic domain walls, in which the electronic states are two dimensional (2D), possibly exist parallel to the surface at the interval between the inverted phase and the original phase (Fig. 1a).
Photon-induced phase inversions were evident in SmTe$_3$ \cite{Trigo2020} from an ultrafast X-ray diffraction experiment and in TbTe$_3$ from an ultrafast optical experiment \cite{Yusupov2010} and also expected in the quasi-one-dimensional CDW material K$_{0.3}$MoO$_3$ based on ultrafast X-ray diffraction data \cite{Huber2014}. However, such photon-induced 2D electronic states in the domain wall have never been identified in a material, and whether the states retain long-range order is still unclear experimentally.

Herein, we report the ultrafast photon-induced long-range 2D ordered electronic states at the surface in a three-dimensional (3D) CDW material. We used infrared ultrashort laser pulses to pump the sample and monitored the electronic structure and lattice dynamics by high-resolution time- and angle-resolved photoemission (trARPES) \cite{Yang2019} and MeV ultrafast electron diffraction (UED) \cite{Qi2020}, respectively (Fig. 1b, see Methods). With improved energy and pump fluence resolution, the trARPES experiments showed evidence of 2D electronic states on the surface due to the ultrafast phase-inversion-induced macroscopic domain wall, which was confirmed by the temporal lattice distortion from high-resolution UED experiments and was consistent with the phenomenological theory based on a spatially and temporally dependent double-wall Ginzburg--Landau potential. Interestingly, a novel phase with enhanced density of states and possible energy gap opening near the Fermi energy, was discovered at the macroscopic domain wall.

\section*{Ultrafast CDW evolution}

The material we studied was  1T-\TS, for which the CDW transition at 202 K (T$_c$) is possibly a result of the cooperation of the excitonic interactions and Jahn--Teller effect \cite{DiSalvo1976,Cercellier2007,MoehrVorobeva2011,Kogar2017}. Below T$_c$, the distortion of two adjacent Se-Ti-Se layers are anti-phase locked, forming a commensurate 3D CDW with a 2 $\times$ 2 $\times$ 2 periodic lattice distortion (PLD) \cite{Hildebrand2018}. Our trARPES experiment at 4 K clearly showed a flattened ``Mexican-hat'' shaped dispersion in the Se 4$p_{x,y}$ band, which is a signature of strong electron--hole interactions (left panel of Fig. 1c) \cite{Pillo2000,Cercellier2007,Watson2019}.
Equilibrium PLD was evident from the diffraction patterns (left panel of Fig. 1d). Furthermore, the Se 4$p_{x,y}$ bands folded from point $A$ ($A$-4$p_{x,y}$) and the Ti 3$d_{z^2}$ band folded from point $L$  ($L$-3$d_{z^2}$) due to the PLD were resolved. At 0.5 ps after the photoexcitation with a pump fluence higher than the breaking threshold of the excitonic correlation and CDW order \cite{Rohwer2011,Porer2014,Hedayat2019}, the flattened Se 4$p_{x,y}$ band could not be well defined near the band top (central panel of Fig. 1c), which was possibly due to the strong CDW fluctuations. The PLD superlattice peaks were almost indistinguishable  (right panel of Fig. 1d) at the same delay time, indicating that both the electronic order and the PLD could be completely suppressed at such a high pump fluence. Both the $L$-3$d_{z^2}$ band and the Se 4$p_{x,y}$ band were restored at a delay time of 12 ps (right panel of Fig. 1c).

To track the ultrafast evolution of the CDW order, we will focus on the dynamics of the flattened Se 4$p_{x,y}$ band at the $\Gamma$ point. At an excitation fluence of 0.02 \mJcm, the time-dependent photoemission spectrum and the PLD diffraction intensity clearly showed a band oscillation in both intensity and energy (Figs. 2a, c, and d),  which was a result of photon-induced coherent $A_{1g}$-CDW phonons \cite{Holy1977,Snow2003,Porer2014,Hedayat2019}. At the same high excitation fluence as that shown in Fig. 1c, the top of the 4$p_{x,y}$ band was not resolvable until 1 ps after photon excitation possibly due to strong order fluctuations at high non-equilibrium electronic temperature (Fig. 2a).
Two critical pump fluences with $F_{c1}$ at 0.073 \mJcm~(quenching of the exciton condensation) (Fig. 2b) and $F_{c2}$ at 0.135 \mJcm~ (suppressing the PLD) (Fig. 2c) were identified (Extended Data Fig. 2). 

In addition, intensity oscillations occur immediately after time zero at high pump fluence values, but detailed analysis showed that the oscillations were anharmonic, and the peak-to-peak distance was shorter than the period of the $A_{1g}$-CDW phonon oscillations (Fig. 2c). Such an anharmonic oscillating state is a signature of a forced vibration, and the measured electronic states even temporally enter inverted CDW states (see discussion in Supplemental Information Section I), which is further confirmed by UED measurements. With increasing pump fluence, more time is required to reach a maximum suppression of the PLD peak, and the coherent oscillations even became invisible by increasing the pump fluence further (blue arrows in Fig. 2d), which is the signature of softening of the $A_{1g}$-CDW phonons. At the highest pump fluence, less time was required to reach the first maximum suppression of the CDW peak, and interestingly, a peak (red arrows in Fig. 2d) reappeared closer to time zero than the first oscillation peak at a low pump fluence, which was a result of pump-induced inhomogeneous PLD inversion in real space.

The above observations indicated the existence of photoinduced phase inversion in the TiSe$_2$ and was consistent with a simulation of the phase inversion scenario based on the phenomenological theory of a spatially and temporally dependent double wall Ginzburg--Landau potential (see Methods and Supplementary Information Section I) \cite{Yusupov2010,Schaefer2014}. 
The solved order parameter and PLD diffraction intensity from the phenomenological theory show similar phenomenon of the anharmonic oscillations (Fig. 2g) and the reappearance of the PLD peak (Fig. 2h) at the highest pump fluence, consistent with the experimental results in Figs. 2c and d.
From the phenomenological theory, the distribution of the domain wall perpendicular to the sample surface can be tuned by the excitation pulse energy (Figs. 2e and f). It can be expected that at certain pump fluences, transient macroscopic domain walls, which are possibly quasi-2D electronic systems, can be positioned near the sample surface (top panels in Fig. 3b) with a thickness of 2-3 unit cells (determined from Fig. 2f).

\section*{Optically induced 2D domain wall}

TrARPES is an excellent tool for probing the transient electronic structure with surface sensitivity. Experimentally, by finely tuning the pump fluence, features of the domain wall should be alternatively shown up in the ARPES data. The detailed pump-fluence-dependent photoemission spectra at the $\Gamma$ point at a delay time of 12 ps showed that the photoemission intensity just slightly increased until a dramatic enhancement occurred above the exciton condensation critical pump fluence $F_{c1}$ (Figs. 3a and b).
Interestingly, the intensity shifted upward reaching a peak at the pump fluence of $\sim$0.091 \mJcm~(I'), and then dropped to a minimum (dip) at $\sim$0.116 \mJcm~(I) (Fig. 3b). Upon enhancing the pump fluence further, the intensity shifted to new maxima (at fluences of 0.156 (II'), 0.237 (III'), and 0.340 \mJcm) and then new minima (at fluences of 0.185 (II), 0.273 (III), and 0.371 \mJcm) alternately until reaching a point at which it monotonically increases at high pump fluences. The difference spectra in Fig. 3c clearly showed reductions of the intensity near the Fermi energy at the $\Gamma$ between fluences I' and I (I$_{E,k}$(I)-I$_{E,k}$(I')), and between fluences II' and II (I$_{E,k}$(II)-I$_{E,k}$(II')). Further experiment showed that the peak-dip feature lasted for longer than 30 ps (Extended Data Fig. 3d).

The above peak-dip features in intensity change is a result of change of band folding from A point, for the A-4$p_{x,y}$ band is slightly above the $\Gamma$-4$p_{x,y}$ band. 
Since the A-4$p_{x,y}$ band was much weaker in intensity and closer to the Fermi energy than the 4$p_{x,y}$-3$d_{z^2}$ band, we attributed the left edge shift (green cut in the inset of Fig. 3d) as a result of the shift of the 4$p_{x,y}$-3$d_{z^2}$ band, which monotonically shifted upward with increasing pump fluence (Fig. 3d). Interestingly, the edge at the bottom right (red cut in the inset of Fig. 3d) of the 4$p_{x,y}$-3$d_{z^2}$ band showed similar peak-dip feature as in the intensity, and the edge at the upper right side (blue cut in the inset of Fig. 3d) of the 4$p_{x,y}$-3$d_{z^2}$ band was similar to that at the bottom right but with a weaker peak--dip feature, which was also attributed to the folded A-4$p_{x,y}$ band. 
The observed dips described above were results of the lower folding intensity from the A-4$p_{x,y}$ band, indicating that the electronic states at the dips with fluences of I, II, and III were possibly 2D. The reduced spectral linewidth at the fluence I and II (Fig. 3d) further proved the 2D electronic states near the surface at those fluences, since due to the effect of momentum resolution perpendicular to the surface, photoemission spectral linewidth from 2D electronic states is narrower than that from 3D electronic states (EDCs in Extended Data Fig. 4e).
Consistent with the calculation based on a CDW inversion induced sharp domain wall (Extended Data Fig. 4), the inset in Fig. 3d and the EDCs in Extended Data Fig. 4\textbf{e} show an increment of the binding energy of Se 4$p_{x,y}$ band on the 2D domain wall (comparison between pump fluence I’ and I).
From the solved order parameter (Fig. 2e), the 2D electronic structure was determined to be on the macroscopic domain wall where the Ginzburg--Landau potential ($V(\Phi)$) was a maximum (red curve in Fig. 3b). 

It is interesting that the intensity of the $L$-3$d_{z^2}$ band showed a minimum (maximum) at the same fluences where the 4$p_{x,y}$ band edge reached a maximum (minimum) (Figs. 3b and c) and can last for 30 ps (Extended Data Figs. 3\textbf{e} and \textbf{f}), indicating that the near Fermi energy density of states of the $L$-3$d_{z^2}$ band was enhanced in the photoinduced 2D electronic states (domain wall). The enhancement of the density of states of the $L$-3$d_{z^2}$ band cannot be a result of loss of excitonic order at higher pump fluence, since it was reduced with further enhanced pump fluence at II’ (the I$_{E,k}$(II')-I$_{E,k}$(I) in Fig. 3c). Such photoinduced 2D electronic states, which result in the peak-dip features in fluence dependent intensity of the 4$p_{x,y}$ band, could also be identified at equilibrium temperatures of 30 and 80 K but were absent at 295 K (Fig. 4a), which is above the CDW transition temperature. 

\section*{Novel 2D electronic states}

Recently, superconductivity was discovered in pressurised, Cu-intercalated, and electric-field controlled 1T-\TS~ \cite{Joe2014,Yan2017,Li2016}. The existence of an incommensurate CDW phase with localised commensurate CDW domains separated by domain walls is common in superconducting 1T-\TS, and the enhancement of the density of states at the domain wall is believed to be responsible for the superconductivity. The 2D domain wall reported herein also showed an enhancement of the density of states near the Fermi energy (the intensity of the $L$-3$d_{z^2}$ band at 0.116 \mJcm~ in Fig. 3b). Such an enhancement of the density of states might also result in superconductivity in 1T-\TS. A close comparison of the energy distribution curves (EDCs) for the pump fluences of 0.091 (peak) and 0.116 \mJcm~(dip) at 4 K showed the signature of leading edge shift, indicating a potential energy gap $\sim$2 meV at the domain wall (Fig. 4c). Potential energy gap ($\sim$1.5 meV) was also evident at the second domain wall on the surface as the pump fluence was increased further (in the EDCs measured at 0.156 (peak) and 0.185 \mJcm~(dip) in Fig. 4c), but no resolvable gap was found away from the domain wall.
Energy gaps were also identified at two other samples and were evidenced up to 30 ps only when the peak-dip features discussed in Fig. 3b were still resolvable (Extended Data Fig. 3\textbf{c} and \textbf{f}). We note that in the superconducting Cu-intercalated 1T-\TS, the estimated energy gap was about 0.5 meV \cite{Li2007}, which was at the same scale as the domain-wall-induced energy gap. Interestingly, the electronic temperature estimated by fitting the Fermi edge to an energy-resolution-convolved Fermi--Dirac function was about 90 K, which is much higher than the equilibrium superconducting transition temperature in the pressurised, Cu-intercalated, and electric-field-controlled samples. 

Potential energy gap was still resolvable ($\sim$1 meV) at 30 K, but absent at 80 K (Fig. 4c). A superconductor-like energy gap with a size comparable to the equilibrium superconducting gap was evident in K$_3$C$_{60}$ from the ultrafast optical spectra after pumping with an ultrashort far-infrared laser pulse at 100 K, which was significantly greater than its equilibrium superconducting transition temperature of 20 K \cite{Mitrano2016}. The potential energy gap observed in the phase-inversion-induced domain wall was similar to the observation for K$_3$C$_{60}$, in which the energy gap was comparable to the equilibrium superconducting gap in spite of its much higher electronic temperature. Such a pump-induced energy gap at a high electronic temperature may share the same non-equilibrium physics, although they may have different mechanism origins. The pioneering observations of light-induced coherent conductivity at high temperatures in cuprates \cite{Fausti2011,Hu2014}, in which the studied materials also possessed high-temperature CDW phases, also possibly originated from the pump-induced macroscopic domain walls. However, we note that the suggestive transient energy gap cannot be an exclusive evidence of superconductivity and could be also resulted from transient CDW, semiconducting, and other novel phases which hold gapped electronic states. Further experimental and theoretical work is required to clarify it.

In summary, we detected photo-induced macroscopic CDW domain walls in 1T-\TS. The found domain walls existed parallel to the surface in the interval between the photo-inverted phase and the original phase and could be placed near the surface by finely tuning the excitation energy. This macroscopic domain wall exhibited the behaviour of a 2D electronic system, and thus, it is a platform for realising novel phases, for example, phases with superconductivity. Our work presents a novel approach for manipulating quantum materials using ultrafast laser pumping, and opens a new window in ultrafast science with profound implications for next-generation devices with new functionalities. However, further studies are necessary to clarify the precise mechanism for producing such macroscopic domain walls, to determine if such methods are universal and applicable for other CDW materials or even other ordered solids, and most interestingly, to identify if the observed energy gap was a result of photoinduced superconductivity.

%\newpage

\section*{Acknowledgments}

W.T.Z. acknowledges support from the Ministry of Science and Technology of China (2016YFA0300501) and from National Natural Science Foundation of China (11974243) and additional support from a Shanghai talent program. Y.F.G. acknowledges the support by the National Natural Science Foundation of China (Grant No. 11874264). D.Q. acknowledges support from the Ministry of Science and Technology of China (Grant No. 2016YFA0301003). D.Q. and W.D.L acknowledges support from the National Natural Science Foundation of China (Grants No. U1632272 and No. 11521404). D.X. and J.Z. acknowledge support from the National Natural Science Foundation of China (Grants No. 11925505, 11504232, and 11721091) and from the office of Science and Technology, Shanghai Municipal Government (No. 16DZ2260200 and 18JC1410700). First-principles computations were performed at the Center for High Performance Computing of Shanghai Jiao Tong University.

\section*{Author Contributions}

S.F.D. and Y.C. contributed equally to this work. W.T.Z. and D.X. proposed and designed the research. S.F.D., Y.Y.Y., C.Z.H., T.W.T., and W.T.Z contributed to the development and maintenance of the trARPES system. S.F.D., Y.Y.Y., and C.Z.H. collected the trARPES data. Y.C. and F.F.Q. took the UED measurements. W.X. and Y.F.G. prepared the single crystal sample. S.F.D., Y.C., D.X., and W.T.Z. did the phenomenological simulation. C.Y.X and W.D.L did the DFT calculation. W.T.Z. wrote the paper with S.F.D., Y.C., D.Q., D.X., and J.Z. All authors discussed the results and commented on the manuscript.

\section*{Competing Interests}
 The authors declare that they have no competing financial interests.

\begin{figure}
\centering\includegraphics[width=1\columnwidth]{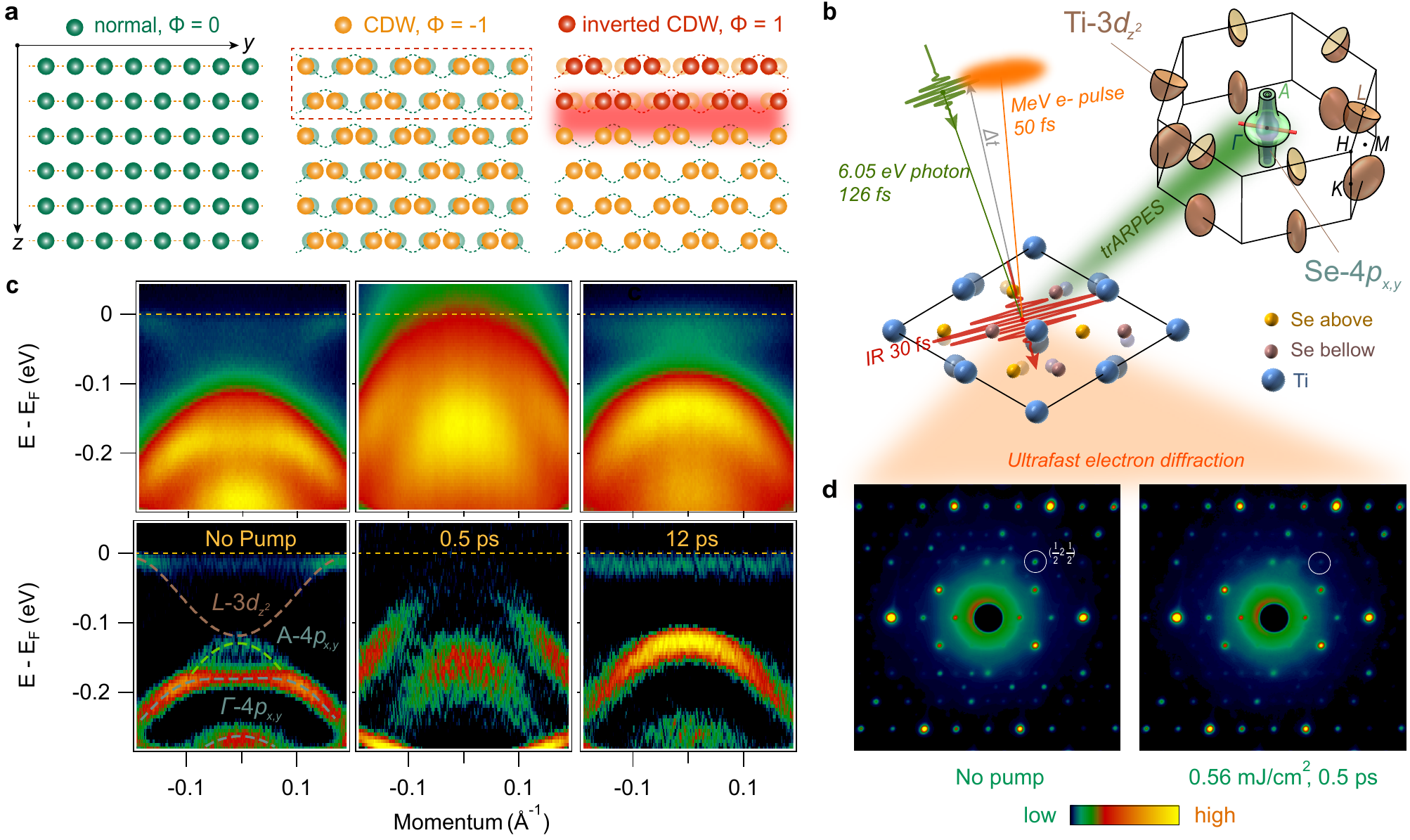}
  \caption{
\textbf{Time-resolved electronic structure and electron diffraction patterns in \TS.}
\textbf{a}, Schematic of the charge density wave (CDW) inversion in real space (side view). The position of the top surface was defined as $z=0$. The normal state lattice was 1 $\times$ 1 $\times$ 1 with the order parameter $\Phi=0$. The CDW lattice was 2 $\times$ 2 $\times$ 2 with the order parameter $\Phi=-1$. The inverted layers were anti-phase distorted with the order parameter $\Phi=1$. Upon photon excitation, phase inversion happens in the red dashed rectangle, and quasi 2D electronic states exist between the inverted and original CDW layers (red shaded area).
\textbf{b}, Experimental geometry of time- and angle-resolved photoemission (trARPES) and ultrafast electron diffraction (UED). $\Delta t$ is the time delay between the infrared pump pulse and ultraviolet laser and MeV electron beams. The momentum cut (red line in the 3D Brillouin zone) in trARPES experiment is centered at $\Gamma$ along $\Gamma - M$ direction.
\textbf{c}, Electronic structures with a pump fluence of 0.355 \mJcm~ and their corresponding 2nd derivative images along energy axis measured at 4 K.
\textbf{d}, Typical UED patterns of \TS~ before and after photoexcitation with a pump fluence of 0.56 \mJcm~ at 90 K. The white circle represents the ($\frac{1}{2}2\frac{1}{2}$) PLD peak.
}
\label{Fig1}
\end{figure}

\begin{figure}
\centering\includegraphics[width=1\columnwidth]{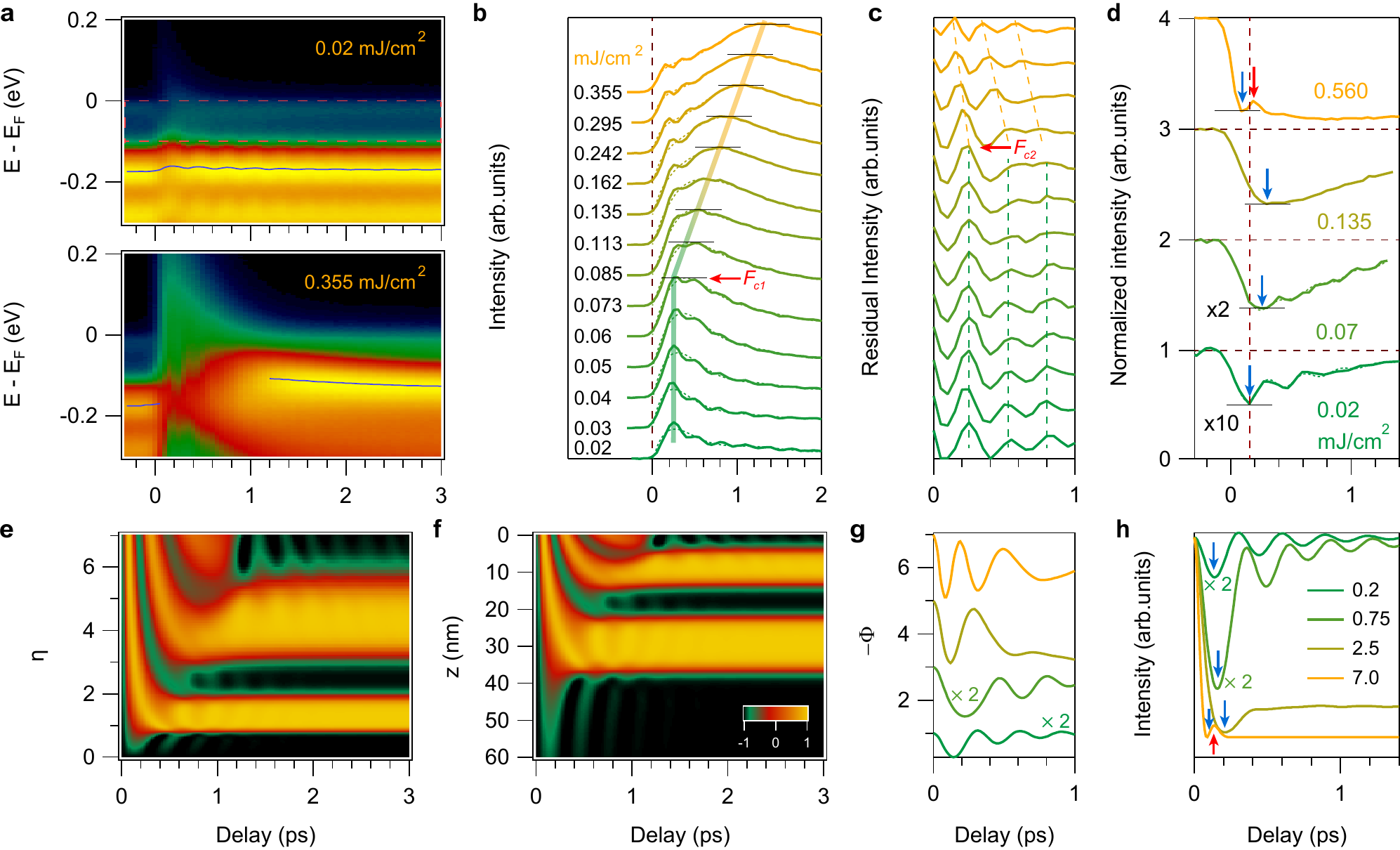}
\caption{
\textbf{Experimental and simulated electronic structure and periodic lattice distortion (PLD) dynamics.}
\textbf{a}, Time-dependent photoemission spectrum at $\Gamma$ for low and high pump fluences. The blue curves are the position of the upper Se 4$p_{x,y}$ band as a function of delay time by tracking the peak position at $\Gamma$.
\textbf{b}, Pump-fluence-dependent spectral intensity as a function of the delay time integrated from $-0.1$ eV (Se 4$p_{x,y}$ band top) to the Fermi level highlighted by the red dashed box in panel \textbf{a}.
\textbf{c}, Residual intensities after removing smooth backgrounds from the curves shown in \textbf{b}. Red arrows denote the two critical pump fluences.
\textbf{d}, Pump-fluence-dependent diffraction intensity as a function of the delay time at the PLD peak ($\frac{1}{2}2\frac{1}{2}$).
\textbf{e} and \textbf{g}, Solution of the order parameter as a function of pump fluence ($\eta$) and delay time on the sample surface ($z=0$).
\textbf{f}, Solution of the order parameter as a function of delay time and depth to the sample surface at a pump fluence ($\eta$) of 7. 
\textbf{h}, Simulated PLD diffraction intensity as a function of delay time at pump fluences ($\eta$) of 0.2, 0.75, 2.5, and 7. In $\textbf{d}$ and $\textbf{h}$, blue arrows denote the delay time when the PLD intensity is mostly suppressed at corresponding pump fluences, and red arrow denotes the reappeared peak after the suppression of the PLD intensity.}

\label{Fig2}
\end{figure}

\begin{figure}
\centering\includegraphics[width=1\columnwidth]{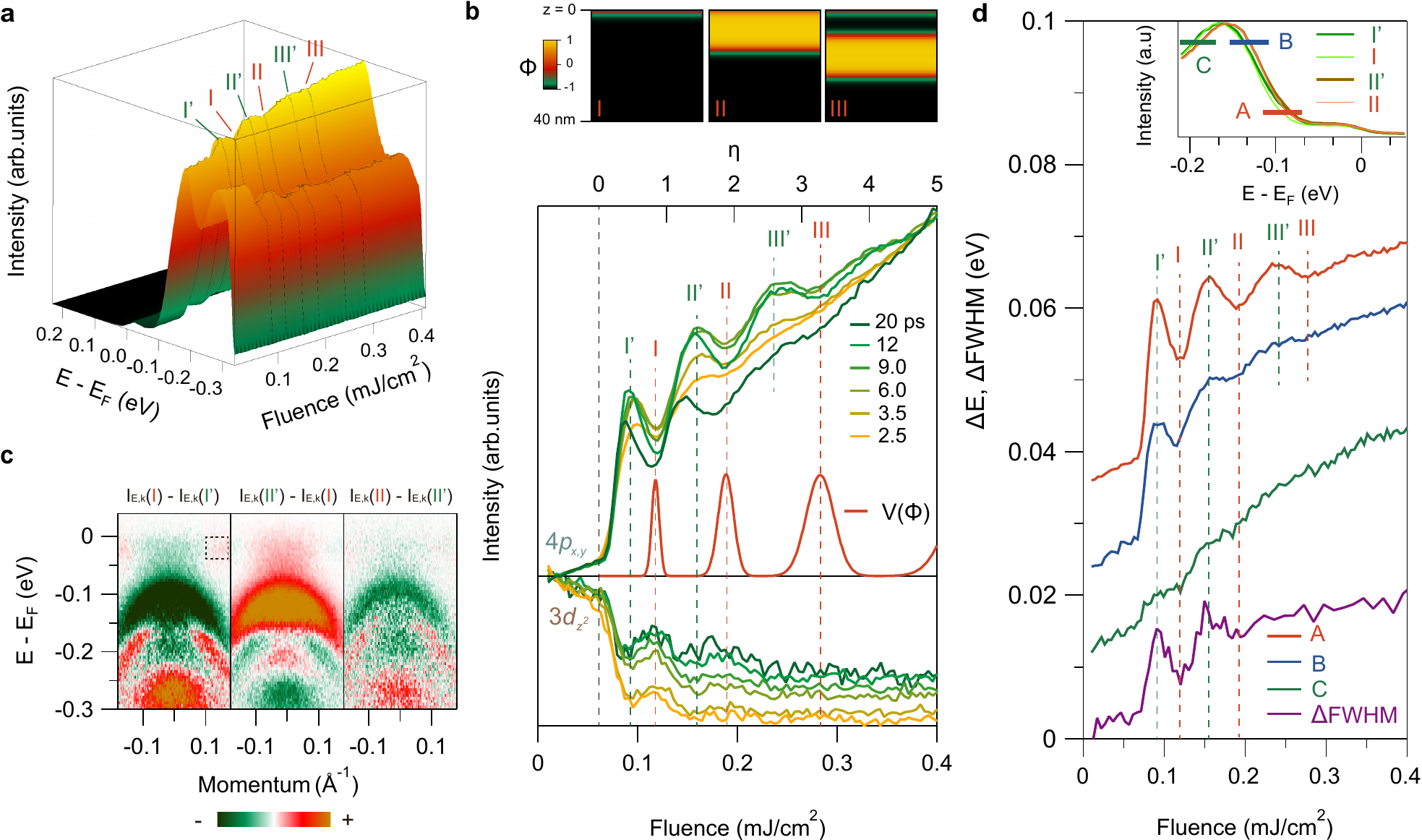}
  \caption{
\textbf{Photoinduced domain wall.}
\textbf{a}, Photoemission spectrum at $\Gamma$ as a function of the pump fluence at 12 ps.
\textbf{b}, Spectral intensity (I$_{E,k}$) as a function of the pump fluence integrated from $-0.1$ eV (Se 4$p_{x,y}$ band top) to the Fermi level and  $L$-3$d_{z^2}$ band integrated in the dashed rectangle shown in the upper left panel of Fig. 3c, calculated $V(\Phi)$ ($\eta$) at the surface ($z=0$). All the intensities are normalized to the same height. Calculated order parameters as a function of depth ($z$) for fluences at I, II, and III are plotted on the top of the panel.
\textbf{c}, Photoemission spectra difference between fluences I' and I (I$_{E,k}$(I)-I$_{E,k}$(I')), I and II' (I$_{E,k}$(II')-I$_{E,k}$(I)), and II' and II (I$_{E,k}$(II)-I$_{E,k}$(II')).
\textbf{d}, Fluence-dependent band edge shifts in energy at the cuts A, B, and C shown in the inset (EDCs for pump fluences at the peaks (0.091 and 0.156 \mJcm) and dips (0.116 and 0.185 \mJcm) in \textbf{a} and \textbf{b}, and the spectral linewidth change ($\Delta$FWHM) of the Se $p_{x,y}$ band at the momentum 0.13 $\AA^{-1}$.
}
\label{Fig3}
\end{figure}

\begin{figure}
\centering\includegraphics{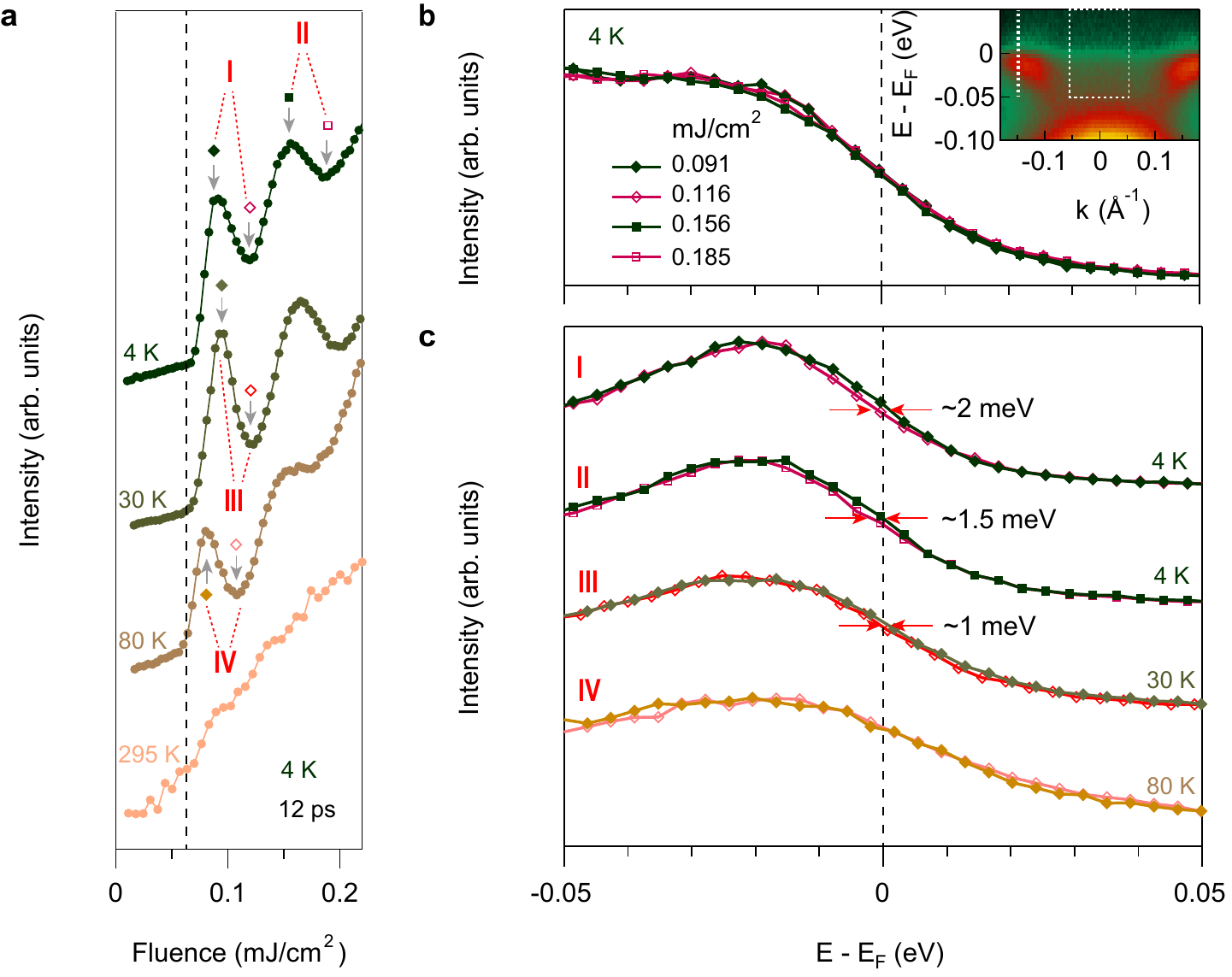}
\caption{
\textbf{Energy gap in the pump-induced domain wall.}
\textbf{a}, Pump-induced photoemission intensity change between $0$ and $-0.1$ eV at $\Gamma$ as a function of the pump fluence at 4, 30, and 80 K and above the CDW transition temperature of 295 K. The measurements were taken 12 ps after the photoexcitation.
\textbf{b}, Integrated photoemission intensity as a function of energy between the momentum values of $-0.05$ and $0.05$ $\AA^{-1}$ (dashed rectangle in the inset) for pump fluences at the peaks and dips (domain walls) at 4 K shown in \textbf{a}. 
\textbf{c}, Energy distribution curves (EDCs) at the momentum of the dashed line shown in the inset of \textbf{b} for pump fluences at the peak--dip features (denoted by I, II, III, and IV in \textbf{a} and \textbf{c}). The red arrows show the energy gap open from the leading edge shift between pump fluences at the peaks and dips at 4 K and 30 K, but no signature of energy gap at 80 K within energy resolution. All the EDCs are normalized to the same height.
}
\label{Fig4}
\end{figure}

\begin{figure}
\centering\includegraphics[width=1\columnwidth]{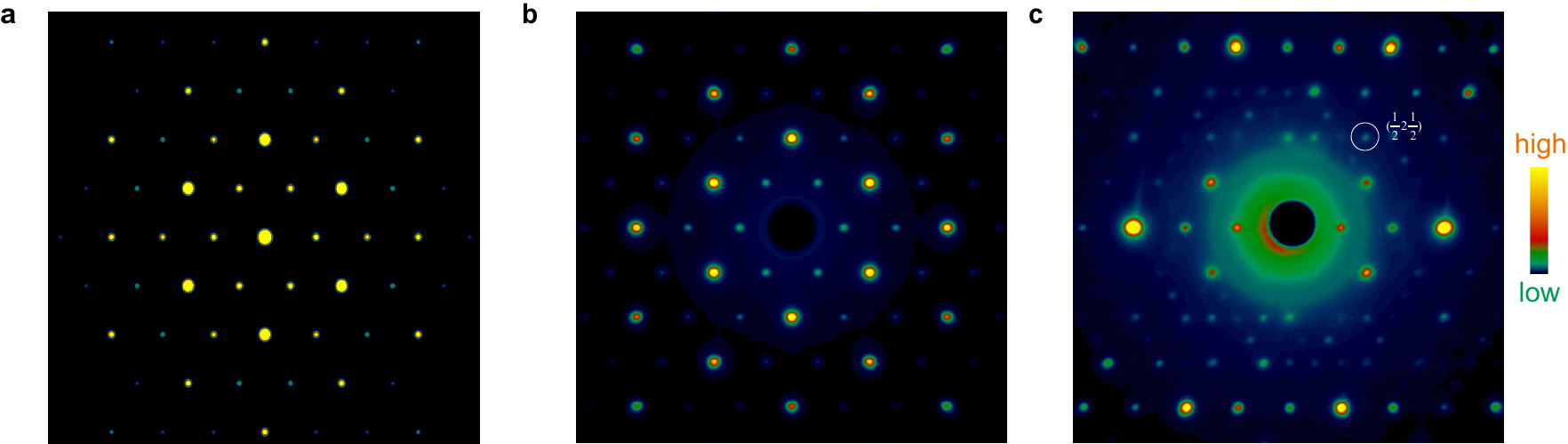}
\caption{
\textbf{Extended Data Fig. 1 Simulated and experimental electron diffraction patterns in \TS.}
\textbf{a} and \textbf{b}, Simulated and experimental electron diffraction patterns based on the kinetical theory with the sample thickness of 60 nm.
\textbf{c},  Experimental diffraction patterns at 12 ps with a tilted angle. Data was taken with the same pump fluence and  equilibrium temperature as that shown in Fig. 1\textbf{d}  of the main text. 
}
\label{eFig1}
\end{figure}

\begin{figure}
\centering\includegraphics{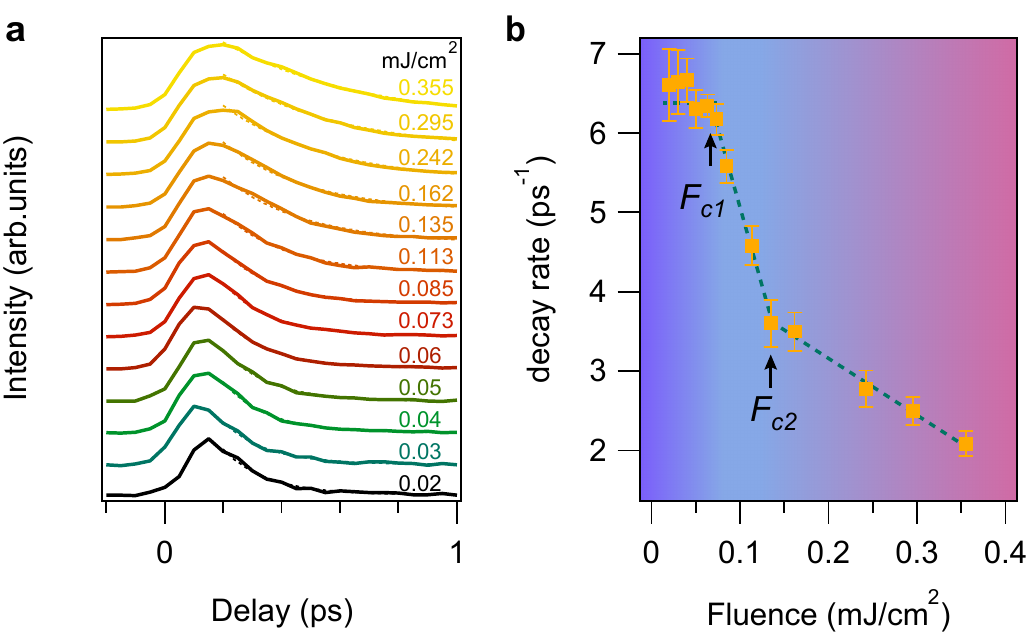}
\caption{
\textbf{Extended Data Fig. 2 Ultrafast electronic dynamics.}
\textbf{a},  Time dependent photoemission spectroscopy intensity at different pump fluences. The intensity is the integration of non-equilibrium electrons between 0 and 0.03 eV above the Fermi energy.
\textbf{b}, The decay rates of nonequilibrium electrons as a function of pump fluence.
}
\label{eFig2}
\end{figure}

 \begin{figure}
\centering\includegraphics[width=1\columnwidth]{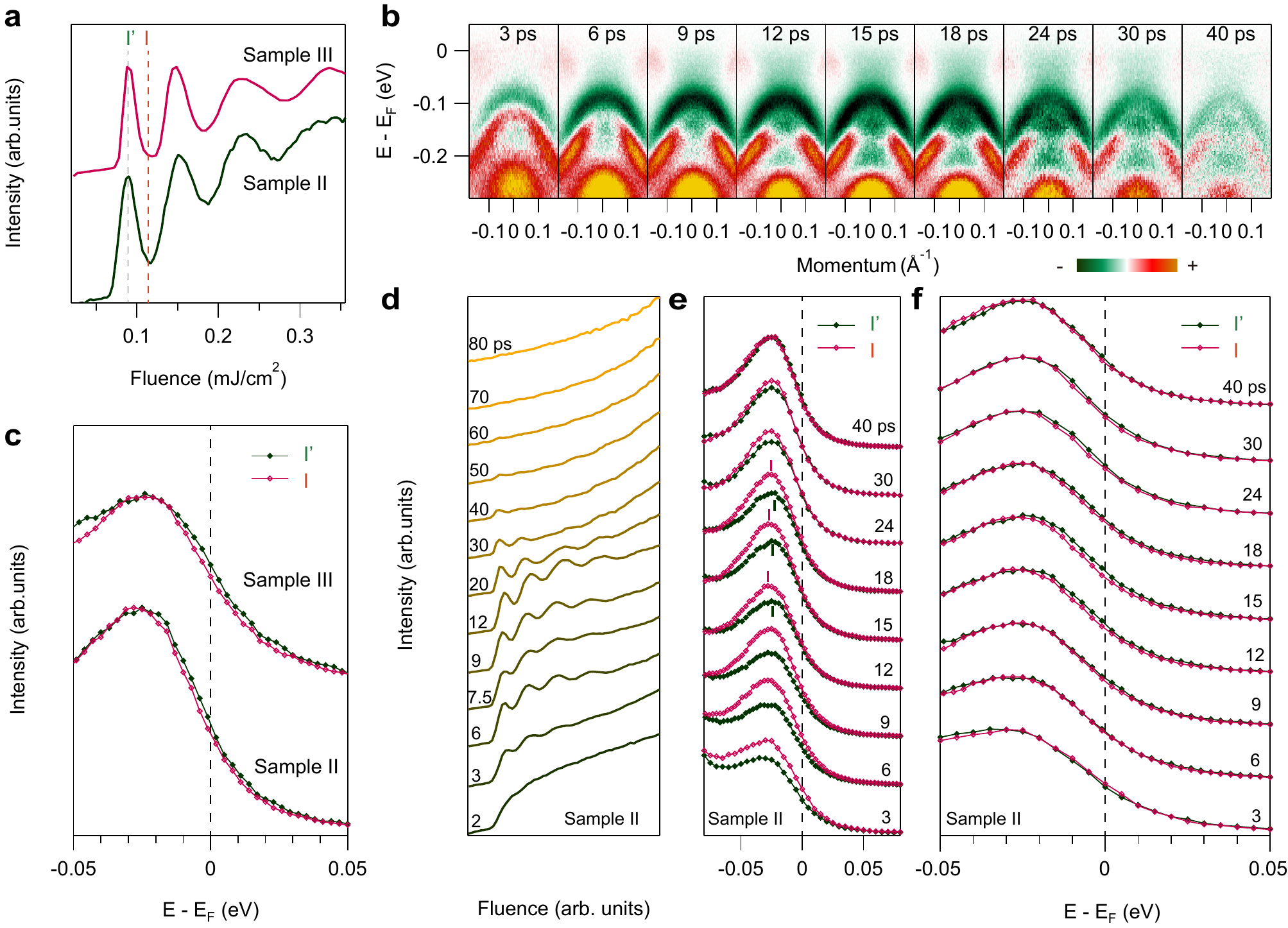}
\caption{
\textbf{Extended Data Fig. 3 Additional measurements on two other samples (Sample II, III).} 
\textbf{a}, Spectral intensity as a function of the pump fluence integrated from $-0.1$ eV (Se 4$p_{x,y}$ band top) to the Fermi level for Sample II and III at the delay time of 12 ps.
\textbf{b} , Photoemission spectra difference between fluences I' and I (I(I)-I(I')) from 3 to 40 ps for sample II.
\textbf{c},  EDCs at the momentum of the dashed line shown in the inset of Fig. 4\textbf{b}  for pump fluences at I' and I for Sample II and III. EDCs are normalized to the same height. 
\textbf{d},  Spectral intensity as a function of the pump fluence integrated from $-0.1$ eV (Se 4$p_{x,y}$ band top) to the Fermi level for Sample II.
\textbf{e},  Original EDCs without normalization between 3 and 40 ps for Sample II.
\textbf{f}  Normalized EDCs from \textbf{e}.
}
\label{eFig3}
\end{figure}

 \begin{figure}
\centering\includegraphics[width=1\columnwidth]{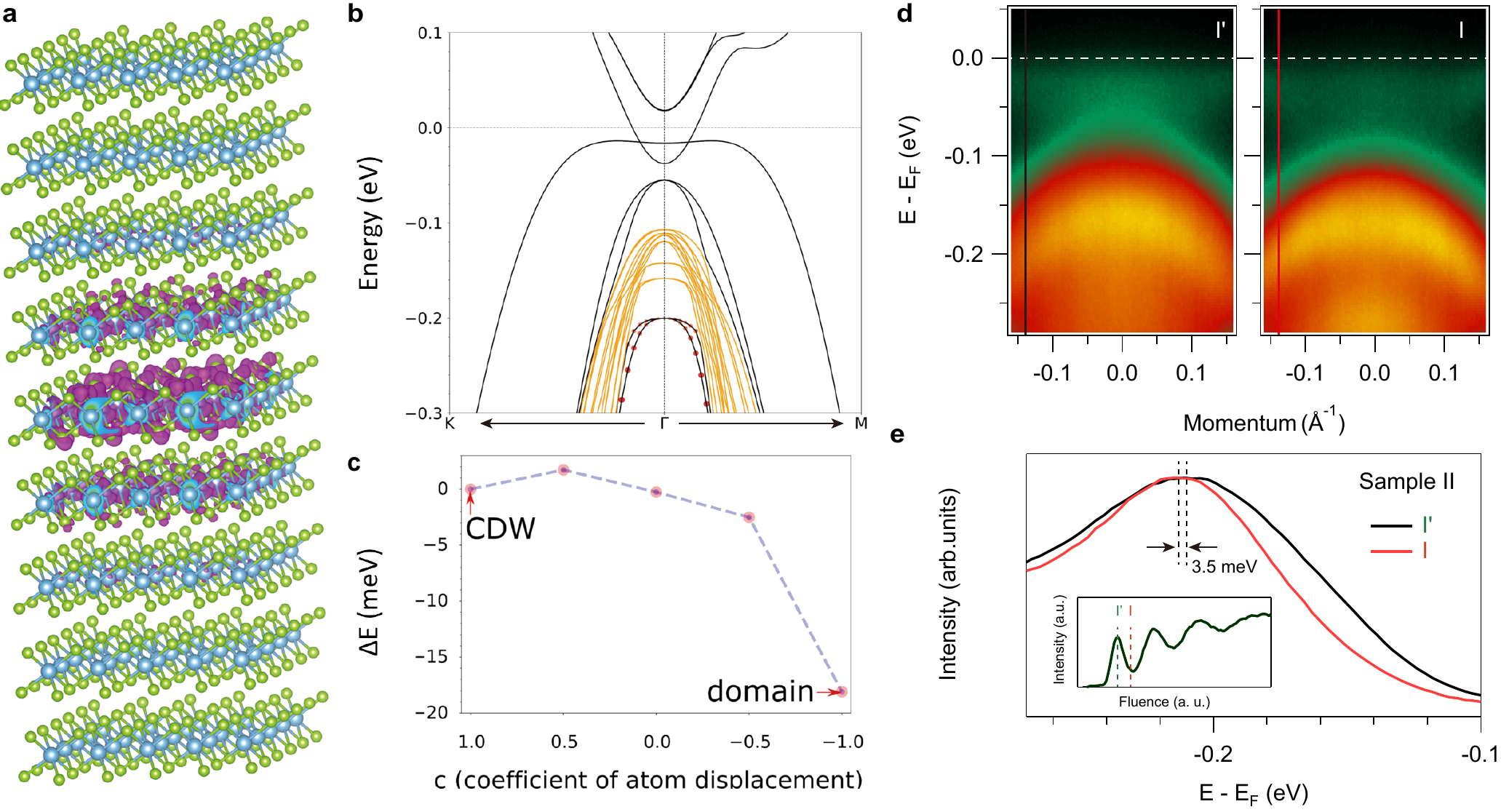}
\caption{
\textbf{Extended Data Fig. 4 DFT calculations and experimental electronic structures of the bulk and 2D domain wall.} 
\textbf{a},  Charge density isosurface plot of the domain-wall bands at $\Gamma$ point. The Ti and Se atoms are shown in blue and green spheres. 
\textbf{b},  Band structure of the periodic 8-layer supercell with domain wall. The $p_x$ and $p_y$ orbitals are projected to the 3 layers at the domain wall (shown in red circles), symbol size denotes the weight of the orbitals. The reference bands are drawn in orange lines. 
\textbf{c},  Energy shifts of the domain-wall bands at different strengths of CDW displacements. The label ``CDW'' means a single CDW phase, while ``domain'' means the presence of sharp domain wall.
\textbf{d},  Time-resolved photoemission spectra at 12 ps for pump fluence at I' and I, as indicated in the inset of the panel \textbf{e} .
\textbf{e},  EDCs at the momentum of -0.13 \AA$^{-1}$ for pump fluence I' and I, as indicated by the solid line cuts (the same colour as the corresponding EDC) in panel \textbf{d} . Inset shows the same spectra at 12 ps for sample II as that shown in Extended Data Fig. 3\textbf{a} . EDCs are normalized to the same height.
}
\label{eFig4}
\end{figure}

\section*{Methods}

\textbf{TrARPES} 

In the trARPES measurements, infrared photon pulses with wavelength centred at 700 nm (1.77 eV)  and pulse length of 30 fs were used to excite the sample, and the non-equilibrium states were probed by ultraviolet pulses at 205 nm (6.05 eV). Both of the pump and probe beam are $s$ polarized in our experimental geometry. The measurements were taken at $k_z=2$ at the Fermi energy ($E_B=0$) by ${k_z} = \frac{1}{{\hbar}}\sqrt {2{m_e}[(hv - W - |{E_B}|){{\cos }^2}\theta  + {V_0}]}$ with photon energy $h\nu$ = 6.05 eV, system work function $W=4.3$ eV, nearly normal emission ($\theta=0$), and inner potential $V_{0}$ = 13.5 eV in our experiments\cite{Watson2019}. The excitation fluences of the pump beam can be precisely tuned by using a half waveplate and a polarizer, and the size of pump pulse is 8.5 times larger than the probe beam, ensuring a fluence resolution of 2.5\% in the fluence dependent measurements (Supplemental Information Section II). Photoelectrons were collected by a Scienta DA30L-8000R analyzer. The overall time resolution and energy resolution are 130 fs and 19 meV respectively \cite{Yang2019}. Data for Sample II and III in the Extended Data Fig. 3 was taken with a time resolution of 112 fs and an energy resolution of 16.3 meV. Sample was cleaved at a pressure better than 3$\times10^{-11}$ torr at 4 K. We note that for 6.05 eV photons, the top 4--5 layers of unit cell can be probed. Limited by the photon energy, we cannot reach the Brillouin zone boundary on our trARPES system. With improved energy and momentum resolution for low energy probe photon, the folded Ti 3d$_{z^2}$ band can be clearly resolved near the Brillouin zone center, making it possible to analysis the dynamics of the Ti 3d band indirectly. 

\textbf{Determining the energy gap}

 In Fig. 4c, we compare EDCs between the peak and dip to extract the energy gap, because at these two corresponding excitation fluences, the electronic temperature was comparable and the temperature effect on the EDC could be excluded. The energy gap here was obtained from the leading edge shift between the EDCs at the two pump fluences, which is a common method for determining the energy gap with a non-apparent coherent peak\cite{Norman1998} and usually underestimates the energy gap. We note that a leading edge shift did not appear in the inelastic scattered electrons, which represented the Fermi--Dirac broadening at these pump fluences (Fig. 4b).

\textbf{MeV UED} 

The structural dynamics was independently investigated by a MeV UED facility with a double bend achromat compressor. An 800 nm laser with about 30 fs pulse width was used to pump the sample and a 3 MeV electron beam is used to probe the dynamics. The electrons were produced in a photocathode radio-frequency gun and were further compressed to below 30 fs (FWHM) at the sample with the compressor \cite{Qi2020}. The size of pump pulse is 5 times larger than the electron beam, thus ensuring uniform excitation on the sample. An electron-multiplying CCD camera and a phosphor screen are used to measure the diffraction pattern. High quality \TS~thin film with a thickness of 60 nm was prepared by exfoliating a bulk single crystal. Sample was tilted by several degree to the normal incidence to avoid the intensity cancellation in the PLD diffraction (Extended Data Fig. 1). During the experiment, the temperature of the sample was kept at $\sim$90 K by liquid nitrogen.

\textbf{Domain wall simulation} 

The pump-induced macroscopic domain wall can be modelled by solving a motion equation based on a spatially- and temporally-dependent double wall Ginzburg-Landau potential \cite{Yusupov2010,Schaefer2014}. The potential $V(\Phi,z)$, in which $z$ is the depth away from the surface and $\Phi$ is the order parameter, is
\begin{equation}
\label{Vphi}
V(\Phi,z)=\frac{1}{4}(\eta\cdot e^{-\frac{t}{\tau_{e}}}\cdot e^{-\frac{z}{z_p}}-1)\Phi^2+\frac{1}{8}\Phi^4+\frac{1}{4}\xi^2(\frac{\partial \Phi}{\partial z})^2.
\end{equation}
Here $\eta e^{-\frac{t}{\tau_{e}}}e^{-\frac{z}{z_p}}$ is the transient change on the potential energy after photoexcitation with the pump fluence coefficient $\eta$ , nonequilibrium quasiparticle recovery rate $\frac{1}{\tau_{e}}=0.3~ps^{-1}$ determined by experiment, and pump penetration depth $z_p=17$ nm \cite{Bayliss1985}, and the CDW coherence length at low temperature $\xi=1.2$ nm \cite{Yusupov2010}. The third term results from inhomogeneous order parameter \cite{McMillan1975a}. The motion euqation based on such a potential is

\begin{equation}
{1 \over {\omega _0^2}}{{{\partial ^2}} \over {\partial {t^2}}}\Phi  + {\gamma  \over {{\omega _0}}}{\partial  \over {\partial t}}\Phi  - {1\over 2}(1 - \eta {e^{ - {t \over {{\tau _{e}}}}}}{e^{ - {z \over {{z_p}}}}})\Phi  + {1\over 2}{\Phi ^3} - {\xi ^2\over2}{{{\partial ^2}} \over {\partial {z^2}}}\Phi  = 0.\label{eq2}
\end{equation}

The temporally and spatially dependent order parameter $\Phi(z,t)$ can be numerically solved from the motion equation with the input of the above parameters in \TS~ (Fig. 2f). The procedure to do the solution is introduced in the Supplementary Information I. 

With the solved $\Phi(z,t)$, the diffraction intensity of the PLD as a function of delay time (Fig. 2h) can be simulated by,
\begin{equation}
I(t) \propto |\int\limits_0^\infty  {\Phi (z,{\rm{t}}){\rm{dz}}} {|^{\rm{2}}}.
\end{equation}

\textbf{UED patterns with normal incidence}

In the CDW state of \TS, the distortion of two adjacent Se-Ti-Se layers are anti-phase locked, forming a commensurate three dimensional charge density wave (CDW) with a periodic lattice distortion (PLD) 2 $\times$ 2 $\times$ 2. Simulation in Extended Data Fig. 1a shows that the first order PLD diffraction peak is zero because of the cancellation between the distorted layer and neighbored anti-phase distorted layer, confirmed by the experimental result shown in Extended Data Fig. 1b. Thus, sample has to be tilted to avoid the cancellation and give the first order of the PLD peaks in the diffraction patterns as shown in the Fig. 1d in the main text.

\textbf{Ultrafast electronic dynamics}

Both the $L$-3$d_{z^2}$ band and the Se 4$p_{x,y}$ band were restored at a delay time of 12 ps (right panel of Fig. 1c in the main text), but with reduced binding energies compared to the equilibrium case and more than half recovered supperlattice diffraction signal (Extended Data Fig. 1c). Although the PLD was almost recovered and the flattening of the Se 4$p_{x,y}$ band was not recovered at 12 ps, it was hard to determine whether the excitonic order was recovered from current data since there might be strongly fluctuated electron-hole interactions.

From the pump-induced change of spectral intensity between the Fermi level and the band top of Se 4$p_{x,y}$ band, it is clear that the photon-induced maximum change of intensity shifts to longer delay time with enhancing the pump pulse energy, showing a critical pump fluence $F_{c1}$ at 0.073 \mJcm~(Fig. 2b in the main text). Besides, the photon-induced coherent $A_{1g}$-CDW phonon persists in the spectrum up to a fluence of 0.135 \mJcm~($F_{c2}$,  Fig. 2c in the main text). 

The two critical pump fluences are also identified in the nonequilibrium quasiparticle recovery rate. Extended Data Fig. 2a shows the time dependent photoemission spectroscopy intensity above the Fermi level at different pump fluences. The recovery rate of the nonequilibrium quasiparticles is obtained by fitting the curves in Extended Data Fig. 2a to an exponential decay function $I(t)=Ce^{-\gamma t}$, in which $\gamma$ is the recovering rate. The obtained recovery rate as a function of pump fluence in Extended Data Fig. 2b clearly shows a critical pump fluence at $F_{c1}$, below which the recovery rate is nearly a constant, and a critical pump fluence at $F_{c2}$, at which there is a kink in the curve.

The observed two critical pump fluences are consistent with that in other reports that the $F_{c1}$ is the fluence that quenching of excitonic order \cite{Porer2014, Hedayat2019} and $F_{c2}$ is the fluence that destroying the period lattice distortion (PLD) \cite{Hedayat2019}.
At fluence below $F_{c1}$, photo carriers are mainly relaxed by $A_{1g}$ phonons, and the recovery time may be mainly governed by the escape rate of the $A_{1g}$ phonon which is supposed to be a constant below $F_{c1}$ from current data.
Above $F_{c1}$, hot carriers enhance Coulomb screening and destroy the excitonic interaction. The coupling between $A_{1g}$ phonons and hot carriers is gradually weakened by enhancing the pump fluence, leading to an increasement of the life time of populated hot carriers above $F_{c1}$. Above $F_{c2}$, PLD and the associated CDW are completely destroyed, showing slower quasiparticle recovery rate due to the temporal uncorrelated states. 
 
 \textbf{The $\textbf{k}_\textbf{z}$ momentum resolution effect}
 
 In ARPES experiments, the final state is confined within the photoelectron escape depth $\lambda$. From the Heisenberg uncertainty principle, such confinement results in intrinsic broadening of $k_z$ (momentum perpendicular to the sample surface) defined by $\Delta k_z\!=\!\lambda^{-1}$, and thus the measured spectrum is an average over the $\Delta k_z$ interval\cite{Strocov2012}. For 3D electronic states, the photoemission spectra are usually broadened due to the $k_z$ resolution, while the photoemission spectra represent the intrinsic 2D electronic states since there is no dispersion along $k_z$ for the 2D case. Due to the $k_z$ resolution, it usually measures sharper ARPES spectra in 2D electronic sates than that in 3D electronic states. Taking the Bi$_2$Se$_3$ for an example, the photoemission spectra for the 2D surface states are much sharper than that of the bulk 3D states, of which the dispersion is even not resolvable\cite{Xia2009}.

For higher pump fluence, laser pulse usually pumps the electronic states to higher electronic temperature and the measured ARPES spectra should be broadened due to enhanced scattering rate at higher electronic temperature. It is quite counterintuitively that with enhancing the optical pump pulse energy the spectral linewidth from photoemission is narrower from the experiment. To our knowledge, it is only the dimension effect that can induce such a phenomenon experimentally. For the above reason, the measured bands getting sharper at higher pump fluence (I) than that at lower fluence (I') is a striking evidence the transient structure is in 2D behaviour. 

\textbf{Additional measurements on two other samples}

The peak-dip feature observed in Fig. 3 was quite reproducible that similar spectral intensities as a function of the pump fluence were taken in two other samples (Extended Data Figs. 3a and d), and was clearly evidenced in sample II until a delay time of 40 ps. Consistent with Figs. 3b and c, the difference of the photoemission spectra between fluence I' and I clearly shows the enhancement of the density of states in the Ti 3$d_{z^2}$ band before 40 ps (Extended Data Figs. 3b and e).

Leading edge shift between fluence I' and I in Fig. 4c is robust and can be evidenced in other samples we measured at the same delay time (Extended Data Fig. 3c), and was evidenced between 9 and 30 ps in Sample II (Extended Data Fig. 3f) and in Sample IV, of which is not shown here. It is clear that the possible energy gap is only shown when there is enhancement of density of states. The absence of energy gap at 3 and 6 ps is possibly due to the high transient electronic temperature close to time 0, and this consistent with Fig. 4c that the energy gap is not resolvable at 80 K although there is still clear peak-dip feature in the fluence dependent measurement, indicating that the electronic temperature must be low enough to present the gapped states.

In addition, the leading-edge shift is a judgement if there was gap opening and usually underestimates the size of the gap for a resolution larger than the gap size. For an example, the measured energy gap on the high-temperature superconductor Bi2212 (overdoped, Tc = 78 K) with an energy resolution of 30 meV from the leading-edge shift is 12 meV\cite{Shen1993}, which is about the half from high resolution data from similar sample\cite{Hashimoto2014}. We note that for comparable spectral linewidth and electronic temperature, it cannot conclude that there is no gap if there is no resolvable leading-edge shift, but on the other hand if there is resolvable leading-edge shift, there must be energy gap although it is hard to determine the exact value. Therefore, the real energy gap in the 2D electronic states on the domain wall should be larger than that obtained from the leading edge shift.

The broadened spectra due to pump excitation and limited energy resolution makes it is hard to determine the energy gap from the quasiparticle peak shift, but it is clear that the peak of the EDCs at the high pump fluence I shifts to higher binding energy at delay time 12, 15, and 18 ps, when the gap is mostly pronounced (Extended Data Fig. 3e).

\textbf{First-principles density-functional-theory calculations}

First-principles density-functional-theory (DFT)\cite{Hohenberg1964,Kohn1965} calculations are performed using the VASP (Vienna Ab initio Simulation Package) code\cite{Kresse1996,Kresse1996a} with the projector augmented wave (PAW) method\cite{Bloechl1994} and the PBE (Perdew-Burke-Ernzerhof) + U approximation\cite{Perdew1997}. A Hubbard interaction parameter U = 3eV on the Ti-$3d$ orbitals is used to obtain the electronic band structure. The lattice constants and displacements of the Ti and Se atoms in the CDW phase are adapted from experimental measurements\cite{DiSalvo1976}. 

In order to understand the electronic structure of the CDW domain wall, calculations are performed on $2\times2\times8$ periodic supercells of TiSe$_2$, in which the atomic displacements from the high-symmetry positions in one particular layer are scaled by a coefficient $c$. A single CDW phase without domain wall corresponds to $c=1$, while $c=0$ means that this one layer has no CDW distortion, and $c=-1$ means this one layer has a reversed CDW distortion, corresponding to sharp CDW domain wall. Through band structure unfolding analysis of the CDW phase, we have identified the $p_x$ and $p_y$ states that originated from the $\Gamma$ point of the high-symmetry phase. In the case of sharp domain wall, the band structure is shown in Extended Data Fig. 4b, with the $p_x$ and $p_y$ domain-wall bands marked by red circles. The spacial distribution of the corresponding charge density is shown in Extended Data Fig. 4a, which exhibits a 2D electronic structure, being extended in the 2D layer while localized in the normal direction.

As the atomic displacements of the domain-wall layer are tuned continuously by the scaling coefficient from $c=1$ to $c=-1$, we calculate the energy shifts of the $p_x$ and $p_y$ domain-wall bands relative to a reference energy. States localized at the domain wall are sensitively affected by the atomic displacements of the domain-wall layer, while delocalized states originated from band-folding in the layer stacking direction are almost unaffected by the domain wall. Thus, these states are good references, and their average energy is used as reference energy for calculating the shift of the domain wall states. As the coefficient $c$ decreases from 1, a phase shift starts to form at the domain-wall layer and the energy of the domain wall states changes, as shown in Extended Data Fig. 4c. In the initial stage, the energy of the domain-wall states remains almost unchanged. When $c=0$, corresponding to a $\pi/2$ phase difference between adjacent layers, the energy shift is about $-0.3$ meV. When the distortion becomes larger at $c=-0.5$ and $1$, the energy shifts become larger. In the case of sharp CDW domain wall ($c=-1$), the energy shift is about $-18$ meV.

In Extended Data Fig. 4d, it clearly shows that the measured electronic structure (on the domain wall) at pump fluence I (dip position in inset of panel e) is sharper than that at pump fluence I' (peak position in inset of panel e). Since the electronic states near the $\Gamma$ are complicated that they are the mixture of A-4p$_{x,y}$, $\Gamma$-4p$_{x,y}$, and the L-3d$_{z^2}$ bands, the pump induced photoemission linewidth and the energy of the 4p$_{x,y}$ band are analysed at the momentum -0.13 \AA$^{-1}$ away from the $\Gamma$. It is clear from Extended Data Fig. 4e that the photoemission spectral linewidth of the Se 4p$_{x,y}$ band is narrower for the pump fluence at I (dip) than that at I' (peak), and the fluence dependence of the spectral linewidth as a function of pump fluence is shown in Fig. 3d in the main text.

It is qualitatively consistent with the DFT calculation that on the domain wall (fluence at I) the Se 4p$_{x,y}$ band moves to higher binding energy than the 3D states (fluence at I'). However, the calculation shows a downshift of the energy band about 18 meV while the experiment shows a downshift of about 3.5 meV as shown in the Extended Data Fig. 4e. The discrepancy between the down shift energy of calculation and experiment is possibly due to the following reasons: 1) a sharp domain wall with only one inverted layer is used in the calculation, while in experiment the change from $\Phi=-1$ to $\Phi=1$ is 2-3 unit cells; 2) the calculation was at zero temperature, while in the measurement the electronic temperature was about 90 K; 3) the downshift energy is estimated from the momentum away from the $\Gamma$ point, at which the energy shift is possibly larger; 4) the DFT calculation might overestimate the energy shift.

The agreement between the calculation and experiment about the energy shift of the band further suggests that the photoinduced electronic states on the domain wall are 2D.

%Method references

\section*{Data Availability}

The data that support the plots within this paper and other findings of this study are available from the corresponding author upon reasonable request. Correspondence and requests for materials should be addressed to W.T.Z. (wentaozhang@sjtu.edu.cn; trARPES experimental data and the simulation) or D.X. (dxiang@sjtu.edu.cn; UED experimental data).
%Extended Figure legends


\begin{thebibliography}{10}
\expandafter\ifx\csname url\endcsname\relax
  \def\url#1{\texttt{#1}}\fi
\expandafter\ifx\csname urlprefix\endcsname\relax\def\urlprefix{URL }\fi
\providecommand{\bibinfo}[2]{#2}
\providecommand{\eprint}[2][]{\url{#2}}

\bibitem{CastroNeto2009}
\bibinfo{author}{Castro~Neto, A.~H.}, \bibinfo{author}{Guinea, F.},
  \bibinfo{author}{Peres, N. M.~R.}, \bibinfo{author}{Novoselov, K.~S.} \&
  \bibinfo{author}{Geim, A.~K.}
\newblock \bibinfo{title}{The electronic properties of graphene}.
\newblock \emph{\bibinfo{journal}{Reviews of Modern Physics}}
  \textbf{\bibinfo{volume}{81}}, \bibinfo{pages}{109--162}
  (\bibinfo{year}{2009}).

\bibitem{Klitzing1980}
\bibinfo{author}{von Klitzing, K.}, \bibinfo{author}{Dorda, G.} \&
  \bibinfo{author}{Pepper, M.}
\newblock \bibinfo{title}{New method for high-accuracy determination of the
  fine-structure constant based on quantized hall resistance}.
\newblock \emph{\bibinfo{journal}{Physical Review Letters}}
  \textbf{\bibinfo{volume}{45}}, \bibinfo{pages}{494--7}
  (\bibinfo{year}{1980}).

\bibitem{Tsui1982}
\bibinfo{author}{Tsui, D.~C.}, \bibinfo{author}{Stormer, H.~L.} \&
  \bibinfo{author}{Gossard, A.~C.}
\newblock \bibinfo{title}{Two-dimensional magnetotransport in the extreme
  quantum limit}.
\newblock \emph{\bibinfo{journal}{Physical Review Letters}}
  \textbf{\bibinfo{volume}{48}}, \bibinfo{pages}{1559--62}
  (\bibinfo{year}{1982}).

\bibitem{He2013}
\bibinfo{author}{He, S.~L.} \emph{et~al.}
\newblock \bibinfo{title}{Phase diagram and electronic indication of
  high-temperature superconductivity at 65 k in single-layer FeSe films}.
\newblock \emph{\bibinfo{journal}{Nature Materials}}
  \textbf{\bibinfo{volume}{12}}, \bibinfo{pages}{605--610}
  (\bibinfo{year}{2013}).

\bibitem{Tan2013}
\bibinfo{author}{Tan, S.~Y.} \emph{et~al.}
\newblock \bibinfo{title}{Interface-induced superconductivity and
  strain-dependent spin density waves in FeSe/SrTiO$_3$ thin films}.
\newblock \emph{\bibinfo{journal}{Nature Materials}}
  \textbf{\bibinfo{volume}{12}}, \bibinfo{pages}{634--640}
  (\bibinfo{year}{2013}).

\bibitem{Yang2019}
\bibinfo{author}{Yang, Y.} \emph{et~al.}
\newblock \bibinfo{title}{{A time- and angle-resolved photoemission
  spectroscopy with probe photon energy up to 6.7 eV}}.
\newblock \emph{\bibinfo{journal}{{Review of Scientific Instruments}}}
  \textbf{\bibinfo{volume}{{90}}}, \bibinfo{pages}{063905} (\bibinfo{year}{{2019}}).

\bibitem{Qi2020}
\bibinfo{author}{Qi, F.~F.} \emph{et~al.}
\newblock \bibinfo{title}{Breaking 50 femtosecond resolution barrier in mev
  ultrafast electron diffraction with a double bend achromat compressor}.
\newblock \emph{\bibinfo{journal}{Physical Review Letters}}
  \textbf{\bibinfo{volume}{124}}, \bibinfo{pages}{134803}
  (\bibinfo{year}{2020}).

\bibitem{Fausti2011}
\bibinfo{author}{Fausti, D.} \emph{et~al.}
\newblock \bibinfo{title}{Light-induced superconductivity in a stripe-ordered
  cuprate}.
\newblock \emph{\bibinfo{journal}{Science}} \textbf{\bibinfo{volume}{331}},
  \bibinfo{pages}{189--191} (\bibinfo{year}{2011}).

\bibitem{Mitrano2016}
\bibinfo{author}{Mitrano, M.} \emph{et~al.}
\newblock \bibinfo{title}{Possible light-induced superconductivity in K$_3$C$_{60}$ at
  high temperature}.
\newblock \emph{\bibinfo{journal}{Nature}} \textbf{\bibinfo{volume}{530}},
  \bibinfo{pages}{461--464} (\bibinfo{year}{2016}).

\bibitem{Ichikawa2011}
\bibinfo{author}{Ichikawa, H.} \emph{et~al.}
\newblock \bibinfo{title}{Transient photoinduced 'hidden' phase in a
  manganite}.
\newblock \emph{\bibinfo{journal}{Nature Materials}}
  \textbf{\bibinfo{volume}{10}}, \bibinfo{pages}{101--105}
  (\bibinfo{year}{2011}).

\bibitem{Stojchevska2014}
\bibinfo{author}{Stojchevska, L.} \emph{et~al.}
\newblock \bibinfo{title}{Ultrafast switching to a stable hidden quantum state
  in an electronic crystal}.
\newblock \emph{\bibinfo{journal}{Science}} \textbf{\bibinfo{volume}{344}},
  \bibinfo{pages}{177--180} (\bibinfo{year}{2014}).

\bibitem{Rini2007}
\bibinfo{author}{Rini, M.} \emph{et~al.}
\newblock \bibinfo{title}{Control of the electronic phase of a manganite by
  mode-selective vibrational excitation}.
\newblock \emph{\bibinfo{journal}{Nature}} \textbf{\bibinfo{volume}{449}},
  \bibinfo{pages}{72--74} (\bibinfo{year}{2007}).

\bibitem{Li2019}
\bibinfo{author}{Li, X.} \emph{et~al.}
\newblock \bibinfo{title}{Terahertz field-induced ferroelectricity in quantum
  paraelectric SrTiO$_3$}.
\newblock \emph{\bibinfo{journal}{Science}} \textbf{\bibinfo{volume}{364}},
  \bibinfo{pages}{1079--1082} (\bibinfo{year}{2019}).

\bibitem{Sie2019}
\bibinfo{author}{Sie, E.~J.} \emph{et~al.}
\newblock \bibinfo{title}{An ultrafast symmetry switch in a weyl semimetal}.
\newblock \emph{\bibinfo{journal}{Nature}} \textbf{\bibinfo{volume}{565}},
  \bibinfo{pages}{61--66} (\bibinfo{year}{2019}).

\bibitem{Horstmann2020}
\bibinfo{author}{Horstmann, J.~G.} \emph{et~al.}
\newblock \bibinfo{title}{Coherent control of a surface structural phase
  transition}.
\newblock \emph{\bibinfo{journal}{Nature}} \textbf{\bibinfo{volume}{583}},
  \bibinfo{pages}{232--236} (\bibinfo{year}{2020}).

\bibitem{Morrison2014}
\bibinfo{author}{Morrison, V.~R.} \emph{et~al.}
\newblock \bibinfo{title}{A photoinduced metal-like phase of monoclinic VO$_2$
  revealed by ultrafast electron diffraction}.
\newblock \emph{\bibinfo{journal}{Science}} \textbf{\bibinfo{volume}{346}},
  \bibinfo{pages}{445--448} (\bibinfo{year}{2014}).

\bibitem{Nova2019}
\bibinfo{author}{Nova, T.~F.}, \bibinfo{author}{Disa, A.~S.},
  \bibinfo{author}{Fechner, M.} \& \bibinfo{author}{Cavalleri, A.}
\newblock \bibinfo{title}{Metastable ferroelectricity in optically strained
  SrTiO$_3$}.
\newblock \emph{\bibinfo{journal}{Science}} \textbf{\bibinfo{volume}{364}},
  \bibinfo{pages}{1075--1079} (\bibinfo{year}{2019}).

\bibitem{Wang2013}
\bibinfo{author}{Wang, Y.~H.}, \bibinfo{author}{Steinberg, H.},
  \bibinfo{author}{Jarillo-Herrero, P.} \& \bibinfo{author}{Gedik, N.}
\newblock \bibinfo{title}{Observation of floquet-bloch states on the surface of
  a topological insulator}.
\newblock \emph{\bibinfo{journal}{Science}} \textbf{\bibinfo{volume}{342}},
  \bibinfo{pages}{453--457} (\bibinfo{year}{2013}).

\bibitem{Lian2020}
\bibinfo{author}{Lian, C.}, \bibinfo{author}{Zhang, S.-J.},
  \bibinfo{author}{Hu, S.-Q.}, \bibinfo{author}{Guan, M.-X.} \&
  \bibinfo{author}{Meng, S.}
\newblock \bibinfo{title}{Ultrafast charge ordering by self-amplified
  exciton-phonon dynamics in TiSe$_2$}.
\newblock \emph{\bibinfo{journal}{Nature Communications}}
  \textbf{\bibinfo{volume}{11}}, \bibinfo{pages}{43} (\bibinfo{year}{2020}).

\bibitem{Trigo2020}
\bibinfo{author}{Trigo, M.} \emph{et~al.}
\newblock \bibinfo{title}{Ultrafast formation of domain walls of a charge density wave in SmTe$_3$}.
\newblock \emph{\bibinfo{journal}{Physical Review B}} 
 \textbf{\bibinfo{volume}{103}}, \bibinfo{pages}{054109}
  (\bibinfo{year}{2021}).
  
  \bibitem{Yusupov2010}
\bibinfo{author}{Yusupov, R.} \emph{et~al.}
\newblock \bibinfo{title}{Coherent dynamics of macroscopic electronic order
  through a symmetry breaking transition}.
\newblock \emph{\bibinfo{journal}{Nature Physics}}
  \textbf{\bibinfo{volume}{6}}, \bibinfo{pages}{681--684}
  (\bibinfo{year}{2010}).


\bibitem{Huber2014}
\bibinfo{author}{Huber, T.} \emph{et~al.}
\newblock \bibinfo{title}{Coherent structural dynamics of a prototypical
  charge-density-wave-to-metal transition}.
\newblock \emph{\bibinfo{journal}{Physical Review Letters}}
  \textbf{\bibinfo{volume}{113}}, \bibinfo{pages}{026401}
  (\bibinfo{year}{2014}).
  


\bibitem{DiSalvo1976}
\bibinfo{author}{Di~Salvo, F.~J.}, \bibinfo{author}{Moncton, D.~E.} \&
  \bibinfo{author}{Waszczak, J.~V.}
\newblock \bibinfo{title}{Electronic properties and superlattice formation in
  the semimetal TiSe$_2$}.
\newblock \emph{\bibinfo{journal}{Physical Review B (Solid State)}}
  \textbf{\bibinfo{volume}{14}}, \bibinfo{pages}{4321--8}
  (\bibinfo{year}{1976}).

\bibitem{Cercellier2007}
\bibinfo{author}{Cercellier, H.} \emph{et~al.}
\newblock \bibinfo{title}{Evidence for an excitonic insulator phase in
  1T-TiSe$_2$}.
\newblock \emph{\bibinfo{journal}{Physical Review Letters}}
  \textbf{\bibinfo{volume}{99}}, \bibinfo{pages}{146403}
  (\bibinfo{year}{2007}).

\bibitem{MoehrVorobeva2011}
\bibinfo{author}{Moehr-Vorobeva, E.} \emph{et~al.}
\newblock \bibinfo{title}{Nonthermal melting of a charge density wave in
  TiSe$_2$}.
\newblock \emph{\bibinfo{journal}{Physical Review Letters}}
  \textbf{\bibinfo{volume}{107}}, \bibinfo{pages}{036403}
  (\bibinfo{year}{2011}).

\bibitem{Kogar2017}
\bibinfo{author}{Kogar, A.} \emph{et~al.}
\newblock \bibinfo{title}{Signatures of exciton condensation in a transition
  metal dichalcogenide}.
\newblock \emph{\bibinfo{journal}{Science}} \textbf{\bibinfo{volume}{358}},
  \bibinfo{pages}{1315--1317} (\bibinfo{year}{2017}).

\bibitem{Hildebrand2018}
\bibinfo{author}{Hildebrand, B.} \emph{et~al.}
\newblock \bibinfo{title}{Local real-space view of the achiral 1T-TiSe$_2$ 2 x 2 x
  2 charge density wave}.
\newblock \emph{\bibinfo{journal}{Physical Review Letters}}
  \textbf{\bibinfo{volume}{120}}, \bibinfo{pages}{136404}
  (\bibinfo{year}{2018}).
  
  \bibitem{Pillo2000}
\bibinfo{author}{Pillo, T.} \emph{et~al.}
\newblock \bibinfo{title}{Photoemission of bands above the fermi level: The
  excitonic insulator phase transition in 1T-TiSe$_2$}.
\newblock \emph{\bibinfo{journal}{Physical Review B}}
  \textbf{\bibinfo{volume}{61}}, \bibinfo{pages}{16213--16222}
  (\bibinfo{year}{2000}).

\bibitem{Watson2019}
\bibinfo{author}{Watson, M.~D.} \emph{et~al.}
\newblock \bibinfo{title}{Orbital-and k(z)-selective hybridization of Se 4p and
  Ti 3d states in the charge density wave phase of TiSe$_2$}.
\newblock \emph{\bibinfo{journal}{Physical Review Letters}}
  \textbf{\bibinfo{volume}{122}}, \bibinfo{pages}{076404}
  (\bibinfo{year}{2019}).

\bibitem{Rohwer2011}
\bibinfo{author}{Rohwer, T.} \emph{et~al.}
\newblock \bibinfo{title}{Collapse of long-range charge order tracked by
  time-resolved photoemission at high momenta}.
\newblock \emph{\bibinfo{journal}{Nature}} \textbf{\bibinfo{volume}{471}},
  \bibinfo{pages}{490--493} (\bibinfo{year}{2011}).

\bibitem{Porer2014}
\bibinfo{author}{Porer, M.} \emph{et~al.}
\newblock \bibinfo{title}{Non-thermal separation of electronic and structural
  orders in a persisting charge density wave}.
\newblock \emph{\bibinfo{journal}{Nature Materials}}
  \textbf{\bibinfo{volume}{13}}, \bibinfo{pages}{857--861}
  (\bibinfo{year}{2014}).

\bibitem{Hedayat2019}
\bibinfo{author}{Hedayat, H.} \emph{et~al.}
\newblock \bibinfo{title}{Excitonic and lattice contributions to the charge
  density wave in 1T-TiSe$_2$ revealed by a phonon bottleneck}.
\newblock \emph{\bibinfo{journal}{Physical Review Research}}
  \textbf{\bibinfo{volume}{1}}, \bibinfo{pages}{023029 (11 pp.)}
  (\bibinfo{year}{2019}).
  
  \bibitem{Schaefer2014}
\bibinfo{author}{Schaefer, H.}, \bibinfo{author}{Kabanov, V.~V.} \&
  \bibinfo{author}{Demsar, J.}
\newblock \bibinfo{title}{Collective modes in quasi-one-dimensional
  charge-density wave systems probed by femtosecond time-resolved optical
  studies}.
\newblock \emph{\bibinfo{journal}{Physical Review B}}
  \textbf{\bibinfo{volume}{89}}, \bibinfo{pages}{045106}
  (\bibinfo{year}{2014}).


\bibitem{Holy1977}
\bibinfo{author}{Holy, J.~A.}, \bibinfo{author}{Woo, K.~C.},
  \bibinfo{author}{Klein, M.~V.} \& \bibinfo{author}{Brown, F.~C.}
\newblock \bibinfo{title}{Raman and infrared studies of superlattice formation
  in TiSe$_2$}.
\newblock \emph{\bibinfo{journal}{Physical Review B (Solid State)}}
  \textbf{\bibinfo{volume}{16}}, \bibinfo{pages}{3628--37}
  (\bibinfo{year}{1977}).

\bibitem{Snow2003}
\bibinfo{author}{Snow, C.~S.}, \bibinfo{author}{Karpus, J.~F.},
  \bibinfo{author}{Cooper, S.~L.}, \bibinfo{author}{Kidd, T.~E.} \&
  \bibinfo{author}{Chiang, T.~C.}
\newblock \bibinfo{title}{Quantum melting of the charge-density-wave state in
  1T-TiSe$_2$}.
\newblock \emph{\bibinfo{journal}{Physical Review Letters}}
  \textbf{\bibinfo{volume}{91}}, \bibinfo{pages}{136402}
  (\bibinfo{year}{2003}).

\bibitem{Joe2014}
\bibinfo{author}{Joe, Y.~I.} \emph{et~al.}
\newblock \bibinfo{title}{Emergence of charge density wave domain walls above
  the superconducting dome in 1T-TiSe$_2$}.
\newblock \emph{\bibinfo{journal}{Nature Physics}}
  \textbf{\bibinfo{volume}{10}}, \bibinfo{pages}{421--425}
  (\bibinfo{year}{2014}).

\bibitem{Yan2017}
\bibinfo{author}{Yan, S.~C.} \emph{et~al.}
\newblock \bibinfo{title}{Influence of domain walls in the incommensurate
  charge density wave state of cu intercalated 1T-TiSe$_2$}.
\newblock \emph{\bibinfo{journal}{Physical Review Letters}}
  \textbf{\bibinfo{volume}{118}}, \bibinfo{pages}{106405}
  (\bibinfo{year}{2017}).

\bibitem{Li2016}
\bibinfo{author}{Li, L.~J.} \emph{et~al.}
\newblock \bibinfo{title}{Controlling many-body states by the electric-field
  effect in a two-dimensional material}.
\newblock \emph{\bibinfo{journal}{Nature}} \textbf{\bibinfo{volume}{529}},
  \bibinfo{pages}{185--189} (\bibinfo{year}{2016}).



\bibitem{Li2007}
\bibinfo{author}{Li, S.~Y.}, \bibinfo{author}{Wu, G.}, \bibinfo{author}{Chen,
  X.~H.} \& \bibinfo{author}{Taillefer, L.}
\newblock \bibinfo{title}{Single-gap s-wave superconductivity near the
  charge-density-wave quantum critical point in Cu$_x$TiSe$_2$}.
\newblock \emph{\bibinfo{journal}{Physical Review Letters}}
  \textbf{\bibinfo{volume}{99}}, \bibinfo{pages}{107001}
  (\bibinfo{year}{2007}).

\bibitem{Hu2014}
\bibinfo{author}{Hu, W.} \emph{et~al.}
\newblock \bibinfo{title}{Optically enhanced coherent transport in YBa$_2$Cu$_3$O$_{6.5}$
  by ultrafast redistribution of interlayer coupling}.
\newblock \emph{\bibinfo{journal}{Nature Materials}}
  \textbf{\bibinfo{volume}{13}}, \bibinfo{pages}{705--711}
  (\bibinfo{year}{2014}).
  
\bibitem{Norman1998}
\bibinfo{author}{Norman, M.~R.} \emph{et~al.}
\newblock \bibinfo{title}{Destruction of the Fermi surface underdoped high-T$_c$
  superconductors}.
\newblock \emph{\bibinfo{journal}{Nature}} \textbf{\bibinfo{volume}{392}},
  \bibinfo{pages}{157--160} (\bibinfo{year}{1998}).

\bibitem{Bayliss1985}
\bibinfo{author}{Bayliss, S.~C.} \& \bibinfo{author}{Liang, W.~Y.}
\newblock \bibinfo{title}{Reflectivity, joint density of states and band
  structure of group IVb transition-metal dichalcogenides}.
\newblock \emph{\bibinfo{journal}{Journal of Physics C (Solid State Physics)}}
  \textbf{\bibinfo{volume}{18}}, \bibinfo{pages}{3327--35}
  (\bibinfo{year}{1985}).

\bibitem{McMillan1975a}
\bibinfo{author}{McMillan, W.~L.}
\newblock \bibinfo{title}{Landau theory of charge-density waves in
  transition-metal dichalcogenides}.
\newblock \emph{\bibinfo{journal}{Physical Review B}}
  \textbf{\bibinfo{volume}{12}}, \bibinfo{pages}{1187--1196}
  (\bibinfo{year}{1975}).
  
  \bibitem{Porer2014}
\bibinfo{author}{Porer, M.} \emph{et~al.}
\newblock \bibinfo{title}{Non-thermal separation of electronic and structural
  orders in a persisting charge density wave}.
\newblock \emph{\bibinfo{journal}{Nature Materials}}
  \textbf{\bibinfo{volume}{13}}, \bibinfo{pages}{857--861}
  (\bibinfo{year}{2014}).
  
  \bibitem{Strocov2012}
\bibinfo{author}{Strocov, V.~N.} \emph{et~al.}
\newblock \bibinfo{title}{Three-dimensional electron realm in
  ${\mathrm{vse}}_{2}$ by soft-x-ray photoelectron spectroscopy: Origin of
  charge-density waves}.
\newblock \emph{\bibinfo{journal}{Physical Review Letters}} \textbf{\bibinfo{volume}{109}},
  \bibinfo{pages}{086401} (\bibinfo{year}{2012}).
  
  \bibitem{Xia2009}
\bibinfo{author}{Xia, Y.} \emph{et~al.}
\newblock \bibinfo{title}{Observation of a large-gap topological-insulator
  class with a single dirac cone on the surface}.
\newblock \emph{\bibinfo{journal}{Nature Physics}}
  \textbf{\bibinfo{volume}{5}}, \bibinfo{pages}{398--402}
  (\bibinfo{year}{2009}).
  
\bibitem{Shen1993}
\bibinfo{author}{Shen, Z.-X.} \emph{et~al.}
\newblock \bibinfo{title}{Anomalously large gap anisotropy in the a-b plane of
  ${\mathrm{Bi}}_{2}$${\mathrm{Sr}}_{2}$${\mathrm{CaCu}}_{2}$${\mathrm{O}}_{8+\mathrm{\ensuremath{\delta}}}$}.
\newblock \emph{\bibinfo{journal}{Physical Review Letters}} \textbf{\bibinfo{volume}{70}},
  \bibinfo{pages}{1553--1556} (\bibinfo{year}{1993}).
  
  \bibitem{Hashimoto2014}
\bibinfo{author}{Hashimoto, M.}, \bibinfo{author}{Vishik, I.~M.},
  \bibinfo{author}{He, R.-H.}, \bibinfo{author}{Devereaux, T.~P.} \&
  \bibinfo{author}{Shen, Z.-X.}
\newblock \bibinfo{title}{Energy gaps in high-transition-temperature cuprate
  superconductors}.
\newblock \emph{\bibinfo{journal}{Nature Physics}}
  \textbf{\bibinfo{volume}{10}}, \bibinfo{pages}{483--495}
  (\bibinfo{year}{2014}).
  
  \bibitem{Hohenberg1964}
\bibinfo{author}{Hohenberg, P.} \& \bibinfo{author}{Kohn, W.}
\newblock \bibinfo{title}{Inhomogeneous electron gas}.
\newblock \emph{\bibinfo{journal}{Physical Review}} \textbf{\bibinfo{volume}{136}},
  \bibinfo{pages}{B864--B871} (\bibinfo{year}{1964}).
  
  \bibitem{Kohn1965}
\bibinfo{author}{Kohn, W.} \& \bibinfo{author}{Sham, L.~J.}
\newblock \bibinfo{title}{Self-consistent equations including exchange and
  correlation effects}.
\newblock \emph{\bibinfo{journal}{Physical Review}} \textbf{\bibinfo{volume}{140}},
  \bibinfo{pages}{A1133--A1138} (\bibinfo{year}{1965}).

\bibitem{Kresse1996}
\bibinfo{author}{Kresse, G.} \& \bibinfo{author}{Furthmuller, J.}
\newblock \bibinfo{title}{Efficiency of ab-initio total energy calculations for
  metals and semiconductors using a plane-wave basis set}.
\newblock \emph{\bibinfo{journal}{Computational Materials Science}}
  \textbf{\bibinfo{volume}{6}}, \bibinfo{pages}{15--50} (\bibinfo{year}{1996}).
  
  \bibitem{Kresse1996a}
\bibinfo{author}{Kresse} \& \bibinfo{author}{Furthmuller}.
\newblock \bibinfo{title}{Efficient iterative schemes for ab initio
  total-energy calculations using a plane-wave basis set.}
\newblock \emph{\bibinfo{journal}{Physical Review B, Condensed matter}}
  \textbf{\bibinfo{volume}{54}}, \bibinfo{pages}{11169--11186}
  (\bibinfo{year}{1996}).

\bibitem{Bloechl1994}
\bibinfo{author}{Bl\"ochl, P.~E.}
\newblock \bibinfo{title}{Projector augmented-wave method}.
\newblock \emph{\bibinfo{journal}{Physical Review B}} \textbf{\bibinfo{volume}{50}},
  \bibinfo{pages}{17953--17979} (\bibinfo{year}{1994}).
  
  \bibitem{Perdew1997}
\bibinfo{author}{Perdew, J.~P.}, \bibinfo{author}{Burke, K.} \&
  \bibinfo{author}{Ernzerhof, M.}
\newblock \bibinfo{title}{Generalized gradient approximation made simple}.
\newblock \emph{\bibinfo{journal}{Physical Review Letters}}
  \textbf{\bibinfo{volume}{78}}, \bibinfo{pages}{1396--1396}
  (\bibinfo{year}{1997}).

\end{thebibliography}
\end{document}

\begin{CJK*}{GBK}{}
\textbf{SUPPLEMENTARY INFORMATION for}
\title{Optical manipulation of electronic dimensionality in a quantum material}
%\title{Possible superconductivity in the photoinduced macroscopic domain wall}
%\title{Possible light-induced superconductivity in the transient domain wall}

\author{Shaofeng Duan}
\thanks{These two authors contributed equally}
\affiliation{Key Laboratory of Artificial Structures and Quantum Control (Ministry of Education), Shenyang National Laboratory for Materials Science, School of Physics and Astronomy, Shanghai Jiao Tong University, Shanghai 200240, China}
\author{Yun Cheng}
\thanks{These two authors contributed equally}
\affiliation{Key Laboratory for Laser Plasmas (Ministry of Education), School of Physics and Astronomy, Shanghai Jiao Tong University, Shanghai 200240, China}
\author{Wei Xia}
\affiliation{School of Physical Science and Technology, ShanghaiTech University, Shanghai 201210, China}
\author{Yuanyuan Yang}
\affiliation{Key Laboratory of Artificial Structures and Quantum Control (Ministry of Education), Shenyang National Laboratory for Materials Science, School of Physics and Astronomy, Shanghai Jiao Tong University, Shanghai 200240, China}
\author{Chengyang Xu}
\affiliation{Key Laboratory of Artificial Structures and Quantum Control (Ministry of Education), School of Physics and Astronomy, Shanghai Jiao Tong University, Shanghai 200240, China}
\author{Fengfeng Qi}
\affiliation{Key Laboratory for Laser Plasmas (Ministry of Education), School of Physics and Astronomy, Shanghai Jiao Tong University, Shanghai 200240, China}
\author{Chaozhi Huang}
\author{Tianwei Tang}
\affiliation{Key Laboratory of Artificial Structures and Quantum Control (Ministry of Education), Shenyang National Laboratory for Materials Science, School of Physics and Astronomy, Shanghai Jiao Tong University, Shanghai 200240, China}
\author{Yanfeng Guo}
\affiliation{School of Physical Science and Technology, ShanghaiTech University, Shanghai 201210, China}

\author{Weidong Luo}
\affiliation{Key Laboratory of Artificial Structures and Quantum Control (Ministry of Education), School of Physics and Astronomy, Shanghai Jiao Tong University, Shanghai 200240, China}
\affiliation{Institute of Natural Sciences, Shanghai Jiao Tong University, Shanghai 200240, China}

\author{Dong Qian}
\affiliation{Key Laboratory of Artificial Structures and Quantum Control (Ministry of Education), Shenyang National Laboratory for Materials Science, School of Physics and Astronomy, Shanghai Jiao Tong University, Shanghai 200240, China}
\author{Dao Xiang}
\email{dxiang@sjtu.edu.cn}
\affiliation{Key Laboratory for Laser Plasmas (Ministry of Education), School of Physics and Astronomy, Shanghai Jiao Tong University, Shanghai 200240, China}
\affiliation{Zhangjiang Institute for Advanced Study, Shanghai Jiao Tong University, Shanghai 200240, China}
\affiliation{Tsung-Dao Lee Institute, Shanghai Jiao Tong University, Shanghai 200240, China}
\author{Jie Zhang}
\affiliation{Key Laboratory for Laser Plasmas (Ministry of Education), School of Physics and Astronomy, Shanghai Jiao Tong University, Shanghai 200240, China}
\author{Wentao Zhang}
\email{wentaozhang@sjtu.edu.cn}
\affiliation{Key Laboratory of Artificial Structures and Quantum Control (Ministry of Education), Shenyang National Laboratory for Materials Science, School of Physics and Astronomy, Shanghai Jiao Tong University, Shanghai 200240, China}

%\date {\today}
\begin{abstract}

\noindent{\textbf{The supplementary information include:}}
\begin{enumerate}[]\bfseries

%  \item \ref{Sec1}. UED patterns with normal incidence
%  \item \ref{Sec2}. Ultrafast electronic dynamics
  \item \ref{Sec3}. Numerical solution of the motion equation
  \item \ref{Sec4}. Pump probe fluence resolution
% \item \ref{Sec5}. The $k_z$ momentum resolution effect
%  \item \ref{Sec6}. Additional measurements on two other samples
%  \item \ref{Sec7}. First-principles density-functional-theory calculations
  
\end{enumerate}

\end{abstract}
\maketitle
\end{CJK*}

\makeatletter
\makeatletter \renewcommand{\fnum@figure}
{\figurename~S\thefigure}
\renewcommand{\theequation}{S\arabic{equation}}
\makeatother
%\begin{equation}
%	R_F=(F_R-F_L) /F_0\cdot 100 \%  
%\end{equation}
%in which $I(F_1)$
%here $x_{left}$ and $x_{right}$ is the position of half photoemission spectroscopy intensity $I\left(x\right)$,  which locate on the left and right sides of the peak respectively, and $FWHM_{pump}$ is the FWHM of pump beam. 
%As we can see pump fluence resolution is much better for higher pump/probe size ratio.

\newpage

\section{Numerical solution of the motion equation}
\label{Sec3}

%As introduced in the methods, the pump induced macroscopic domain wall at the surface can be modeled by solving a motion equation based on a spatially- and temporally- dependent double wall Ginzburg-Landau potential \cite{Yusupov2010}. The potential V($\Phi$,z), in which z is the depth away from
%the sample surface and $\Phi$ is the order parameter, is
%\begin{small}
%\begin{equation}
%V(\Phi ,z) = {1 \over 4}(\eta {e^{ - {t \over {{\tau _{disp}}}}}}{e^{ - {z \over {{z_p}}}}}-1){\Phi ^2} + {1 \over 8}{\Phi ^4} + {1 \over 4}{\xi ^2}{({{\partial \Phi } \over {\partial z}})^2}.\label{eq1}
%\end{equation}
%\end{small}
%Here $\eta {e^{ - {t \over {{\tau _{disp}}}}}}{e^{ - {z \over {{z_p}}}}}-1$ is the transient change on the potential energy after photoexcitation with the pump fluence coefficient $\eta$, ${{\tau _{disp}}}=0.3$ ps is the nonequilibrium quasiparticle recovery time which can be determined by the decay rate in Fig. S\ref{Fig2}b. $z_p=17$ nm is the pump laser penetration depth \cite{Bayliss1985} and $\xi=1.2$ nm \cite{Yusupov2010} is the CDW coherence length at low temperature. The third term results from inhomogeneous order parameter \cite{McMillan1975a}.

We numberically solved the motion equation (2) in the methods following the procedure:
%\begin{small}
%\begin{tiny}
%\begin{footnotesize}
\begin{scriptsize}
\begin{equation}
%\resizebox{0.9}{!}
%{
\left( \begin{gathered}t\!=\!0\\ \Phi \left(z, 0\right)  \!=\!-1\\ {\partial \Phi(z,t) \over\partial t}\!=\!0 \\
	{{{\partial ^2\Phi(z,t)}} \over {\partial {z^2}}} \!=\! 0 \end{gathered} \right)  
	\!\rightarrow\!
\left( \begin{gathered}t\!=\!\delta t\\ \Phi \left(z, t\right)  \!=\!\Phi \left(z, 0\right)\\ {\partial \Phi(z,t) \over\partial t}\!=\!{\partial \Phi(z,t) \over\partial t} \mid_{t=0}\\
	{{{\partial ^2\Phi(z,t)}} \over {\partial {z^2}}} \!=\! {{{\partial ^2\Phi(z,t)}} \over {\partial {z^2}}}\mid_{t=0} \end{gathered} \right)  
	\!\rightarrow\!
\left( \begin{gathered}t\!=\!2\delta t\\ \Phi \left(z,t\right)  \!=\!\Phi \left(z,\delta t\right)\\ {\partial \Phi(z,t) \over\partial t}\!=\!{\partial \Phi(z,t) \over\partial t} \mid_{t=\delta t}\\
	{{{\partial ^2\Phi(z,t)}} \over {\partial {z^2}}} \!=\! {{{\partial ^2\Phi(z,t)}} \over {\partial {z^2}}} \mid_{t=\delta t} \end{gathered} \right)  
	\!\cdots\!
\left( \begin{gathered}t\!=\!n\delta t\\ \Phi \left(z, t\right)  \!=\!\Phi \left(z,(n-1)\delta t\right)\\ {\partial \Phi(z,t) \over\partial t}\!=\!{\partial \Phi(z, t) \over\partial t}\mid_{t=(n-1)\delta t} \\
	{{{\partial ^2\Phi(z, t)}} \over {\partial {z^2}}} \!=\! {{{\partial ^2\Phi(z,t)}} \over {\partial {z^2}}} \mid_{t=(n-1)\delta t}  \end{gathered} \right). 
%	}	
\end{equation}
%\end{footnotesize}
\end{scriptsize}
%\end{tiny}
%\end{small}

1) An initial condition ($\Phi(z,0)\!=\!-1$, $\partial \Phi(z, t) \over\partial t$$=\!0$, and $\partial^2 \Phi(z, t) \over\partial z^2$$=\!0$) is given at $t\!=\!0$;

2) The motion equation at each $z$ is solved with the initial condition in 1), and the $\Phi(z,\delta t)$, $\partial \Phi(z, t) \over\partial t$$\mid_{t=\delta t}$, $\partial^2 \Phi(z, t) \over\partial z^2$$\mid_{t=\delta t}$ can be calculated from the solved equation;

3) The $\Phi(z,2\delta t)$, $\partial\Phi(z, t) \over\partial t$$\mid_{t=2\delta t}$, $\partial^2\Phi(z, t) \over\partial t^2$$\mid_{t=2\delta t}$, can be calculated by solving the motion equation with the initial condition $\Phi(z,\delta t)$, $\partial \Phi(z, t) \over\partial t$$\mid_{t=\delta t}$,$\partial^2\Phi(z,t)\over\partial z^2$$\mid_{t=\delta t}$ in step 2);

4) The above procedure is executed until the end of the delay time ($n\delta t$);

5) After finished the above procedure, based on the solved $\Phi(z,t)$ and the calculated ($\partial \Phi(z, t) \over\partial t$, $\partial^2 \Phi(z, t) \over\partial^2 z$ )  from the solved $\Phi(z,t)$ are as the input to run the procedure step by step again;

6) Usually the solution will be stable after executing step 5) for 2-3 times. The precision of the solution is dependent on the time delay step size $\delta t$.

During the ultrafast process, laser pulses knock down the CDW order without disturbing PLD through inducing a homogeneous bonding-antibonding electronic transition. The reduced CDW order then triggers the ions to move coherently antiparallel to the original PLD. At very low pump fluence ($\eta \to 0$), the Ginzburg-Landau potential is nearly time independent, and the displacements of the ions are small and those ions oscillate near the equilibrium sites at low temperature acted as the photoinduced A$_{1g}$ CDW amplitude mode observed in ultrafast experiment (Figs. 2c, d, g, and h). At high pump fluence, the displacements of the ions are large enough to enter its normal state or even inverted position, as schematized in Fig. 1a, and now the potential $V(\Phi, z)$ is time dependent (the size of the displacement of  the ions), giving the motion of the ions in an anharmonic way that the displacement of the ions is still in the same phase but without certain frequency, creating macroscopic inverted states. There is possibly some incoherent (random phase) component after pumping for fluence above F$_{c2}$ in real material, but for the case of \TS here, the incoherent component is small at fluence not far above F$_{c2}$. With increasing the pump fluence, the incoherent component is possibly enhanced, confirmed by the fact that the domain wall effect is hard to resolve at high pump fluence. 

At the very high pump fluence, the phase inversion happens in a very short time scale ($\sim$100 fs, Figs. 2e and f), anharmonic oscillation in the order parameter (Fig. 2g) and leading to the tiny peak rising after the photoexcitation in calculated UED diffraction intensity (Fig. 2h), which qualitatively agree with the experimental observation in Fig. 2 in the main text.
We note that in real space, the phase inversion ($\Phi\!=\!1$) is corresponding to a lattice distortion with atoms moving towards the opposite direction of the original CDW phase (left panel of Fig.1a in the main text). The consistence between simulations and experimental results demonstrates the feasibility of the phenomenological model in describing the ultrafast evolution of both the electronic state and lattice after photon excitation.

\begin{figure*}[htb]
\centering\includegraphics[width=0.5\columnwidth]{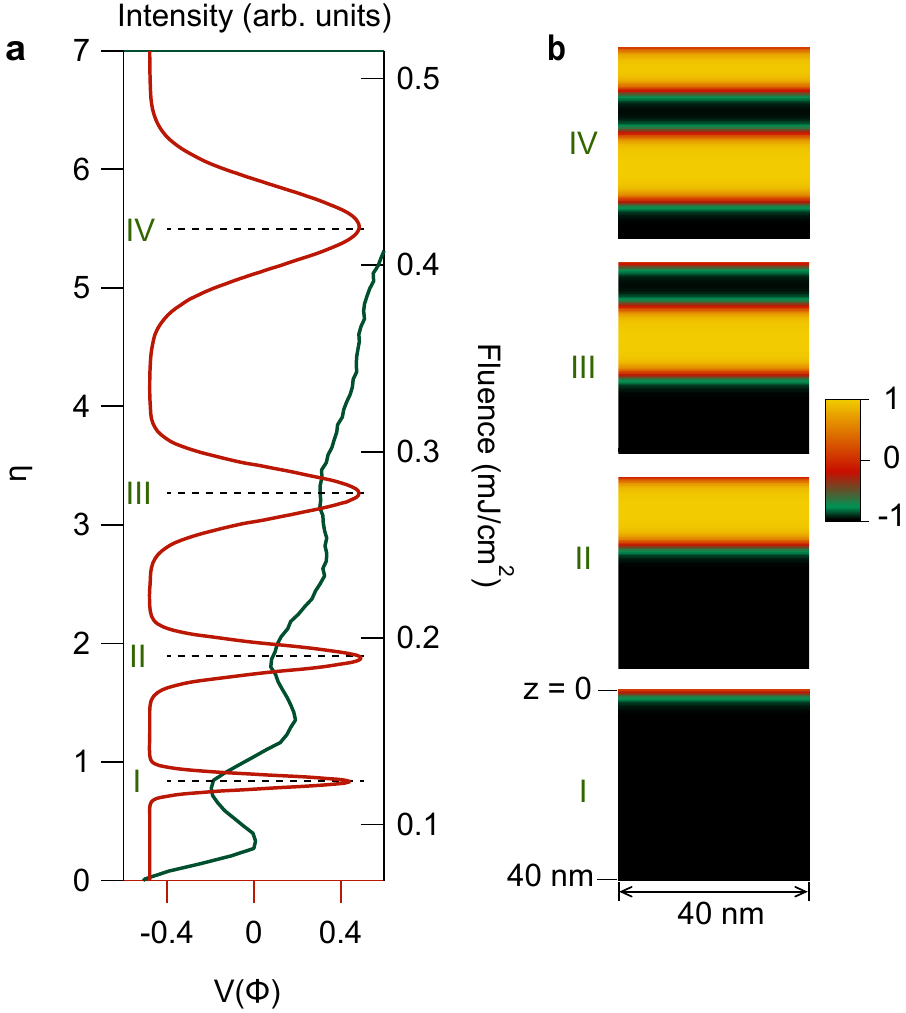}
\caption{
\textbf{Domain walls modeled by a spatially- and temporally-dependent double wall Ginzburg-Landau potential.}
%(a) and (b), Solution of the order parameter as a function of pump fluence and delay time on the sample surface ($z=0$). Order parameter at time zero is -1.
%(c)  Solution of the order parameter as a function of delay time and depth to the sample surface at a pump fluence of 7. 
%(d) Simulated PLD diffraction intensity as a function of delay time at pump fluences of 0.2, 0.75, 2.5, and 7.
(a) Calculated Ginzburg-Landau potential as a function of pump fluence at the surface at 3 ps.  Spectral intensity as a function of pump fluence at 12 ps from Fig. 3b in the main text is plotted for comparison. The start of the right axis is 0.07 \mJcm.
(b) Distributions of the order parameter inside the sample for four different fluences (denoted by I, II, III, IV in (a)), of which the excitation domain wall is right on the surface. $z\!=\!0$ is at the sample surface.
}
\label{FigS3}
\end{figure*}

%Based on the above Ginzburg-Landau potential Eq.~\ref{eq1}, the motion equation is \cite{Yusupov2010},
The solved order parameter from the motion equation as a function of depth $z$ away from the surface at the highest pump fluence clearly shows four domain walls, which are at the interval between the inverted phase ($\Phi\!=\!1$) and the original phase ($\Phi\!=\!-1$), at delay time longer than 1 ps after the photoexcitation (Fig. 2f). Calculations show that the Ginzburg-Landau potential is peaked at the domain walls in the bulk. By finely tuning the pump fluence, the domain wall can be placed at the surface to establish a quasi 2D electronic system. As shown in the simulated order parameter of 3 ps at the surface (Fig. 2e), surface domain walls are expected for $\eta\!=\!$ 0.8, 1.9, 3.2, and 5.5, and the Ginzburg-Landau potential is maximum at those fluences (Fig. S\ref{FigS3}a). The distribution of the simulated order parameter inside the sample shows that the domain wall can be precisely tuned to sit at the surface at the corresponding pump fluences that the Ginzburg-Landau potential is maximum at $z\!=\!0$  (Fig. S\ref{FigS3}b). Interestingly, it has to subtract a value of about 0.07 \mJcm~ from the experimental pump fluence to make the fluences consistent with the calculated ones, at which the pump pulse induces domain wall at $z\!=\!0$, noting that less folded A-4$p_{x,y}$ intensities (dips in the green curve in Fig. S\ref{FigS3}a) is the character that laser pump induces domain wall at the surface.  
By comparing the experimental result to the calculation, $V(\Phi)$ must start near $F_{c1}$ to match the experimental result, which means that the phase inversion was strongly correlated with the Jahn--Teller PLD and the order parameter in the Ginzburg--Landau potential does not need to include the contribution from the excitonic interactions necessarily.

The schematic of phase inversion that a jump in the $\Phi$ value taking place only within 2 CDW layers in Fig. 1a of the main text is a case of very short CDW coherent length. We note that the PLD intensity enhanced at the delay time noted by the red arrows in the Figs. 2d and h is the delay time  when the whole bulk of the material is nearly inverted.

\section{Pump probe fluence resolution}
\label{Sec4}

\begin{figure*}[h]
\centering\includegraphics[width=1\columnwidth]{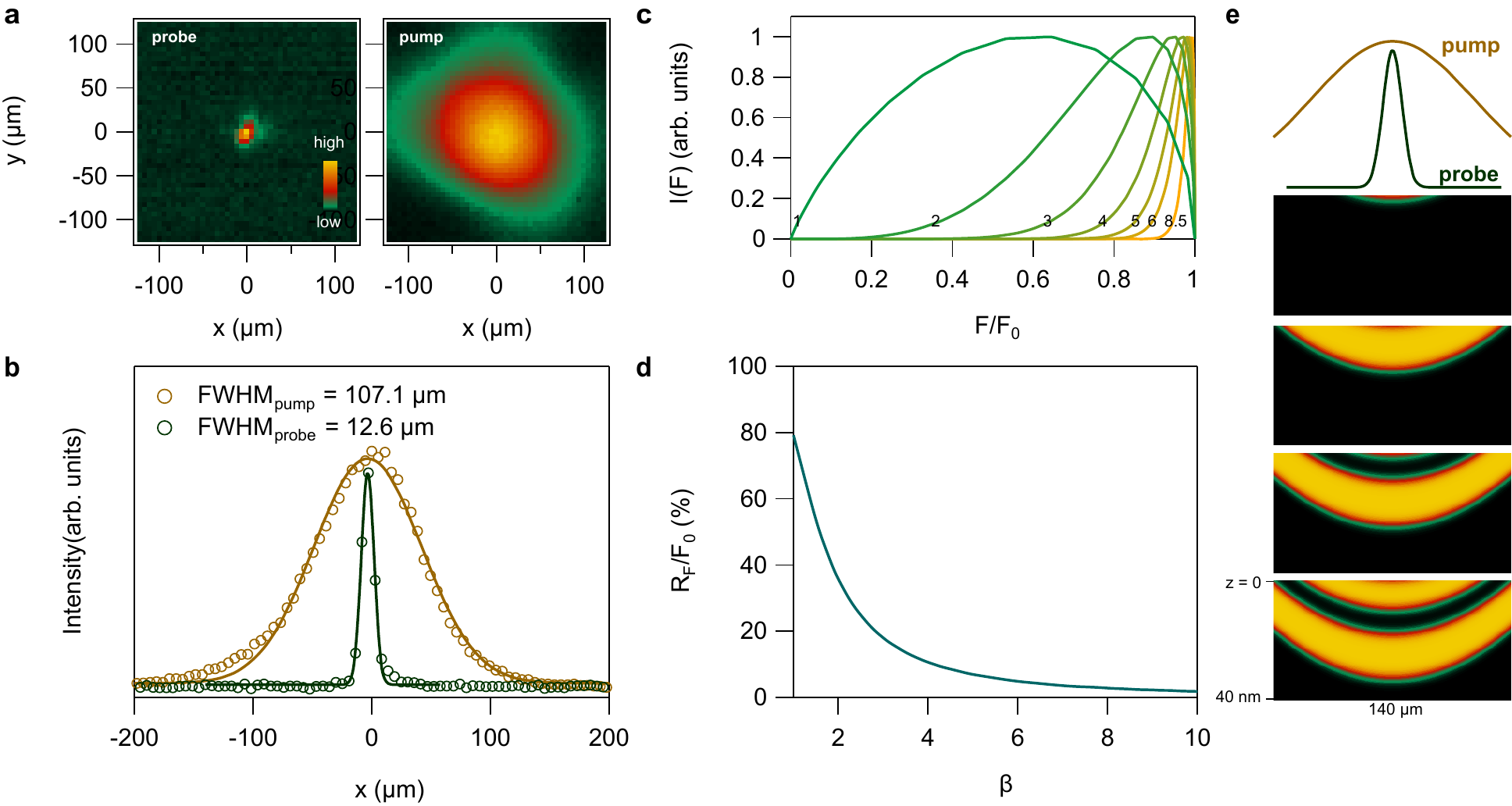}
\caption{
\textbf{Pump probe fluence resolution.} 
(a) Typical beam profiles of probe (205 nm) and pump (700 nm) in our trARPES experiment.
(b) The pump and probe intensities along x axis from a.
(c) Simulated photoemission spectroscopy intensity as a function of pump fluence at different pump/probe size ratio.
(d) Simulated pump fluence resolution as a function of pump/probe size ratio.
(e) The simulated domain walls with a pump of FWHM = 107 $\mu$m.
} 
\label{FigS4}
\end{figure*}

%Since the distribution of the domain wall is strongly dependent on the pump fluence, the probe beam size must be small enough for a Gaussian pump beam.

For the first peak and dip feature observed at 4 K, the corresponding pump fluence are 0.090 (peak) and 0.115 (dip) \mJcm~ respectively (Fig. 4a in the main text), between which the fluence is increased by 28\%. Thus, to detect the photoinduced domain wall in trARPES, pump fluence resolution should be good enough in the fluence dependent  measurements shown in Figs. 3 and 4 in the main text.

The beam profiles of the probe and pump are shown in Fig. S\ref{FigS4}a, and the obtained FWHM of the pump and probe beams are 107.1 and 12.6 $\mu$m respectively, which are the best pump/probe beam size ratio used in our trARPES experiments, and the relative pump fluence is about 2.5\% as discussed in the following.

Since the distribution of the domain wall is dependent on the pump fluence, we simulate the photoemission spectroscopy intensity as a function of  pump fluence to discuss the effect of pump fluence and pump/probe beam size ratio on the distribution of photoemission intensity. The intensity distribution of pump beam can be described by a Gaussian function,
\begin{equation}
	I_{pump}(r) \propto e^{-\frac{r^{2}}{FWHM_{pump}^{2}} \cdot 4ln2}
	%I_{probe}(r) = e^{-\frac{r^{2}}{FWHM_{probe}^{2}} \cdot 4ln(2)}
\end{equation}
The pump fluence $F$ as a function of distance from the beam center $r$ is,
\begin{equation}
	F(r)= F_0\cdot e^{-\frac{r^{2}}{FWHM_{pump}^{2}} \cdot 4ln(2)} \propto I_{pump}(r)
	%I_{probe}(r) = e^{-\frac{r^{2}}{FWHM_{probe}^{2}} \cdot 4ln(2)}
\end{equation}
Here $F_0$ is the fluence at $r=0$. The distribution of the probe beam is,
\begin{equation}
	%I_{pump}(r) = e^{-\frac{r^{2}}{FWHM_{pump}^{2}} \cdot 4ln(2)}\\
	I_{probe}(r) = Be^{-\frac{r^{2}}{FWHM_{probe}^{2}} \cdot 4ln2}
\end{equation}
Then the photoemission intensity $I(F)$ as a function of pump fluence is,
\begin{equation}
	I(F) = 2\pi\cdot r \cdot I_{probe}(r)=2\pi\cdot B\cdot FWHM_{pump}\cdot(\frac{F}{F_0})^{\beta^2}\cdot\sqrt{\frac{ln(F_0)-ln(F)}{4ln2}}
\end{equation}
Here $\beta=\frac{FWHM_{pump}}{FWHM_{probe}}$ is pump/probe beam size ratio. Fig. S\ref{FigS4}c shows the $I(F)$ for $\beta$=1, 2, 3, 4, 5, 6, and 8.5. We define the FWHM of $I(F)$ as the pump fluence resolution ($R_F$). The obtained relative pump fluence resolution $R_F/F_0$ as a function pump/probe beam size ratio $\beta$ is ploted in Fig.~S\ref{FigS4}d. 

For the best fluence resolution we use in our trARPES experiments, the pump/probe beam size ratio $\beta=8.5$ with $FWHM_{pump}=107.1$ and $FWHM_{probe}=12.6$ $\mu$m, and the corresponding relative pump fluence resolution is about 2.5\%. The fluence resolution is 0.0025 \mJcm~ taking the averaging pump fluence 0.1 \mJcm~ for an example. The simulated domain wall distributions for the pump fluences 0.116, 0.185, 0.273, and 0.371 \mJcm, at which a domain wall is right at the surface is plotted in Fig.~S\ref{FigS4}e for a Gaussian pump beam with $FWHM_{pump}=107.1$ $\mu$m. We can see that the probe beam must be small enough to probe the quasi 2D area induced by the phase inversion.

%\cite{DiSalvo1976}. 

%\bibliography{TiSe2bydsf}